\documentclass[]{tJHR2e}

\usepackage{epstopdf}
\usepackage{subfigure}
\usepackage{color}
\usepackage[longnamesfirst,sort]{natbib}
\bibpunct[, ]{(}{)}{;}{a}{,}{,}
\usepackage{siunitx}
\usepackage{graphicx}

\usepackage{xcolor}

\usepackage[colorlinks=true,citecolor=blue,linkcolor=blue]{hyperref}
\usepackage[normalem]{ulem}
\usepackage{lineno}
\newcommand{\f}{\mathtt{F}}

\newcommand{\Reg}{\mbox{\text{Re}}_\text{g}}

\newcommand{\Fr}{\mbox{\text{Fr}}_{\text{air}}}

\newcommand{\rd}{\sigma}
\newcommand{\q}{\mbox{$\hat{\text{q}}_{\text{air}}$}}

\renewcommand{\vec}[1]{\textbf{\textit{#1}}}
\usepackage{cleveref}
\crefformat{figure}{(#2#1#3)}

\begin{document}

\articletype{Guide}
\articletype{Research paper}

\title{Lock-exchange flow regimes under low air Froude number bubble curtains}

\author{Shravan K.R. Raaghav, \textit{Fluid Dynamics Laboratory and J.M. Burgers Centre, Department of Applied Physics, Eindhoven University of Technology, PO Box 513, 5600 MB, Eindhoven, The Netherlands}\\
\textit{Email: s.r.kaveripuram.ramasamy@tue.nl}
\and
Herman J.H. Clercx, \textit{Fluid Dynamics Laboratory and J.M. Burgers Centre, Department of Applied Physics, Eindhoven University of Technology, PO Box 513, 5600 MB, Eindhoven, The Netherlands}\\
\textit{Email: h.j.h.clercx@tue.nl}
\and
Matias Duran-Matute, \textit{Fluid Dynamics Laboratory and J.M. Burgers Centre, Department of Applied Physics, Eindhoven University of Technology, PO Box 513, 5600 MB, Eindhoven, The Netherlands}\\
\textit{Email: m.duran.matute@tue.nl (author for correspondence)}}

\received{v2.1 released September 2025}

\maketitle

\noindent Regimes under low Froude air bubble curtains

\newpage

\noindent Lock-exchange flow regimes under low air Froude number bubble curtains

\begin{abstract}
The flow and density field characteristics around a bubble curtain in a laboratory scale lock-exchange setup are investigated using two-phase large-eddy simulations. We study the detailed hydrodynamics and show that there are three qualitatively distinct (sub)regimes within the previously classified breakthrough regime. The occurrence of these regimes depends not only on air Froude number that characterises the relative strength of the bubble curtain and the gravity current, but also on an additional non-dimensional parameter: the density ratio between the salt and fresh water. The dependence on this additional parameter is also observed in how effective bubble curtains are in blocking the transport of salt to the fresh part of the lock. Hence, it has important implications for the optimisation of bubble curtains in ship locks. 
\end{abstract}


\section{Introduction}
Bubble plumes are multiphase buoyant plumes that have long been of relevance for a wide range of engineering applications \citep{socolofsky2002multi, brevik2002flow, boufadel2020review, dissanayake2018integral}. They are commonly used in bubble columns and air-lift reactors due to their ability to enhance mixing and mass transfer \citep{chisti1987airlift, kantarci2005bubble}. When individual bubble plumes are placed on a line they form a bubble curtain or screen, which can enhance mixing or act as a barrier between two fluids with different properties. As a result, bubble curtains have gained a lot of interest as a solution to various environmental flow problems. For example, they can be used for underwater noise mitigation \citep{wursig2000development}, lake destratification/aeration \citep{wen1987aeration, schladow1993lake, li2025numerical, murai2025density, bombardelli2007modeling}, barriers trapping oil and sediments \citep{mcclimans2013pneumatic, liu2025research, cutroneo2014check, wang2024particle, dugue2015influencing, covarrubias2025interaction}, controlling ichthyoplankton dispersion \citep{prasad2026controlling}, and preventing intrusion of sea water in rivers \citep{nakai2002experimental} and in ship locks \citep{abraham1973pneumatic,keetels2011field,oldeman2020numerical, bacot2022bubble,o2024effect}. The latter is the focus of this paper. 

When the gate of the ship lock that separates the dense seawater and the fresh river water is opened for the ships to pass through, due to the baroclinic pressure gradient, a lock-exchange flow occurs. In this flow, the dense salt water flows beneath the fresh water as a gravity or density current \citep{benjamin1968gravity, shin2004gravity}, causing the intrusion of salt water into the rivers \citep{van1968prediction}. With increasing sea level and severity of droughts, the problem of salt intrusion demands more measures to mitigate salinisation of fresh water reserves \citep{lee2024increasing, Li2025}. Bubble curtains are a promising choice for mitigation because the vertical momentum of the bubbly flow can effectively impede this infiltration of saltwater. In fact, bubble curtains are already installed and working in several ship locks in the Netherlands and more are being planned. However, several questions remain about the fluid dynamics of bubble curtains to optimize their use.

The earliest known studies on the topic are the seminal works of \citet{abraham1962reduction} and \citet{abraham1973pneumatic} who laid down the fundamental theory and concepts behind the hydrodynamics and performance of bubble curtains in lock-exchange configurations by means of theoretical modelling and field measurements. They showed that the amount of saltwater that infiltrates the bubble curtain depends mainly on the air Froude number ($\Fr$), which describes the relative strength of the bubble curtain to the gravity current.

\begin{figure}
    \centering
    \includegraphics[scale=0.4]{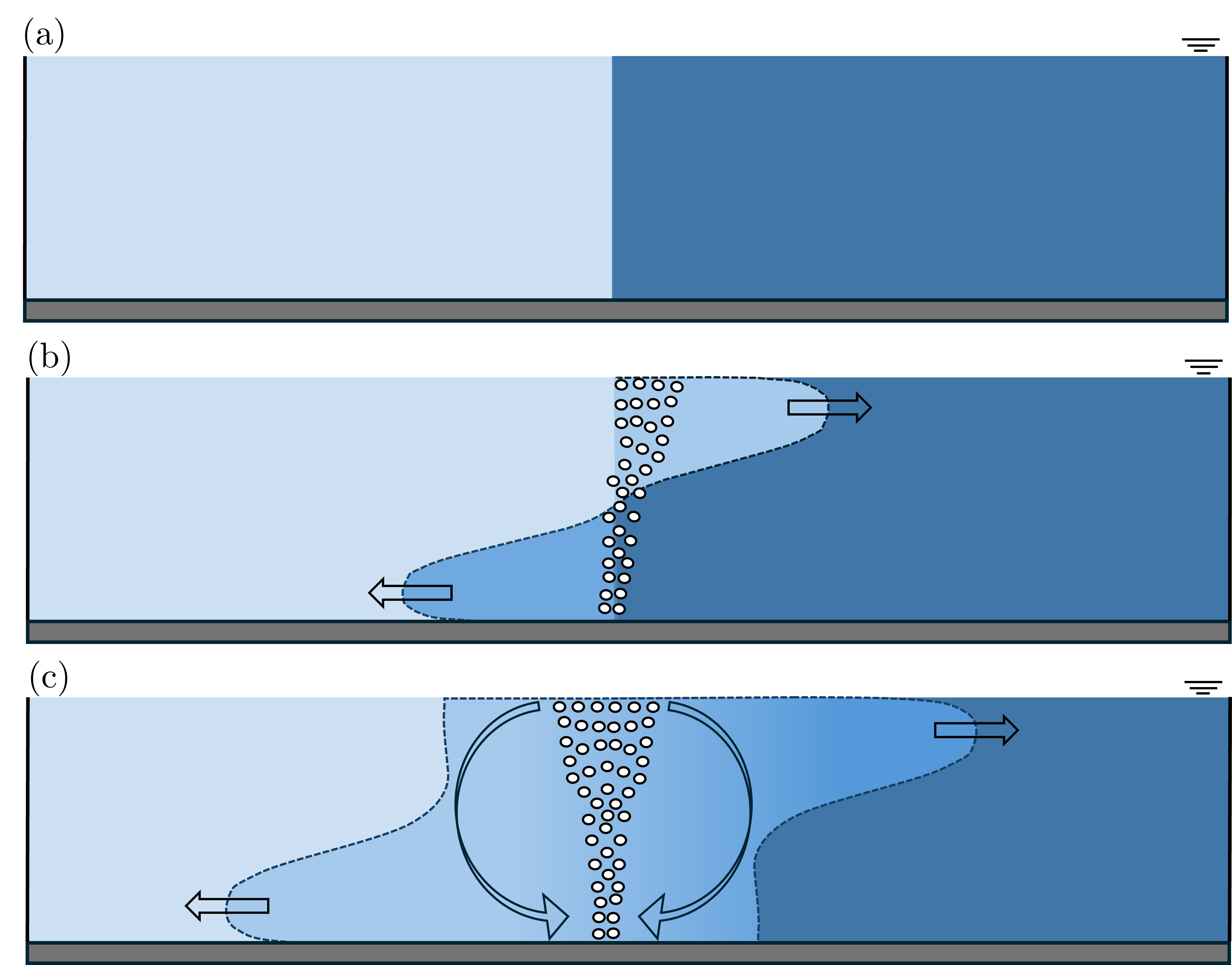}
    \caption{Sketch of the initial lock-exchange configuration (a), with saltwater coloured in dark blue and freshwater coloured with light blue. The two regimes classified by \citet{bacot2022bubble} are shown in b) the breakthrough regime and c) the curtain-driven regime. The horizontal arrows depict the flow of (secondary) gravity currents and the curved arrows depict the recirculation cells formed in the curtain-driven regime.}
    \label{fig:regimes_introduction}
\end{figure}

Interest in this topic has seen a resurgence over the past 15 years. With advances in optical measurement techniques and computational fluid dynamics (CFD), various studies have contributed to an improved fundamental understanding of the hydrodynamics of bubble curtains in a lock-exchange flow. \citet{keetels2011field} combined field-scale experiments, laboratory experiments and numerical simulations to investigate the performance of bubble curtains and in combination with water jets and rigged sills. \citet{van2018methods} conducted particle image velocimetry (PIV) measurements and dye measurements, which were later used for the validation of numerical simulations by \citet{oldeman2020numerical}. They performed Euler-Lagrange large-eddy simulations and were the first to capture an optimal $\Fr$ value ($\Fr\approx0.9$) for maximum effectiveness, where the effectiveness quantifies the amount of salt water that is blocked when a bubble curtain is present compared to the case where it is not. This optimal value was later experimentally confirmed by \citet{bacot2022bubble}, who drew an analogy to air curtains and proposed two regimes based on the optimal $\Fr$ value (Figure \ref{fig:regimes_introduction} shows sketches of the two regimes). For $\Fr\lesssim 0.9$, the breakthrough regime is observed in which the curtain does not possess enough inertia to stop the gravity current. As a result, the gravity current breaks through the curtain (see Figure \ref{fig:regimes_introduction}b). For $\Fr\gtrsim 0.9$, the curtain-driven regime is observed, where the curtain entrains fluid on both sides, forming recirculation cells, from which diluted secondary gravity currents emerge (see Figure \ref{fig:regimes_introduction}c). In addition, \citet{bacot2022bubble} developed a semi-analytical model in the curtain-driven regime using steady-state mass conservation to compute the effectiveness. \citet{raaghav2025bubble} showed, by means of numerical simulations and an unsteady semi-analytical model, that the time elapsed after the opening of the lock gate is also an important parameter to determine the effectiveness of a bubble curtain in the curtain-driven regime. \citet{o2024effect} investigated the effect of the bubble size on flow characteristics and effectiveness. They reported that larger bubbles are more effective in keeping the salt and fresh halves separate and observed that the optimal value of $\Fr$ increased for larger bubbles. They attributed this dependence to the influence of bubble size on entrainment by the bubble curtain.

Most recent studies have focused on obtaining an overview of the dependence of the effectiveness of bubble curtains as a function of $\Fr$ over a wide range of values \citep[see e.g.][]{keetels2011field,oldeman2020numerical,bacot2022bubble} or have focused on a detailed understanding of the curtain-driven regime \citep[see e.g.][]{bacot2022bubble,raaghav2025bubble}. However, for practical purposes, 
operating the bubble curtain close to the optimum but in the breakthrough regime is preferable to operating it in the curtain-driven regime because less energy is needed to achieve similar effectiveness. Hence, this paper focuses on the breakthrough regime. It provides, by means of Euler-Euler large eddy simulations, a detailed characterisation of the flow and density fields as a function of the governing parameters. This characterisation reveals that there are, in fact, three qualitatively distinct sub-regimes, and that the transition between them depends not only on $\Fr$ but also on an additional non-dimensional parameter: the density ratio between the salt and fresh water. The dependence on this additional parameter is also observed in the effectiveness of the bubble curtain, and hence it has important implications for the optimisation of bubble curtains in ship locks.

The remainder of this paper is organised as follows. In Section \ref{section:setup}, we describe the problem, including the relevant non-dimensional parameters. This is followed by the description of the numerical methodology in Section \ref{section:numerical_methodology}. Section \ref{section:regimes_qualitative_description} presents qualitative descriptions of the sub-regimes, and a quantitative characterisation is presented in Section \ref{section:quantitative_description}. A discussion of the relevance and implications of the results for practical applications is given in Section \ref{section:discussion}. Finally, the conclusions are summarised in Section \ref{section:conclusions}.

\section{Problem setup and definition of the parameters}
\label{section:setup}
\begin{figure}
    \centering
    \includegraphics[scale=1]{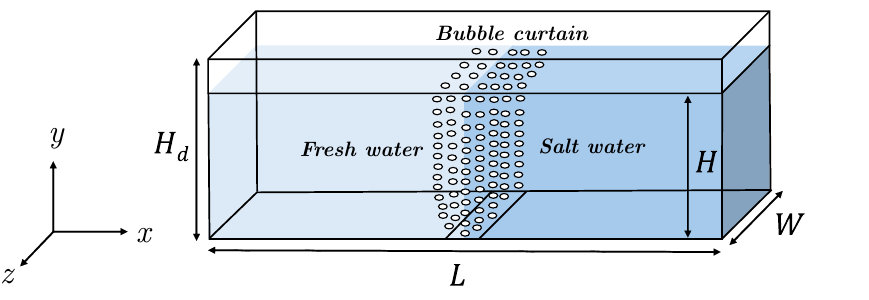}
    \caption{Schematic of a laboratory scale lock with a bubble curtain placed at the centre. The annotations $L$, $W$, and $H_d$ depict the length, width and height of the tank, respectively. The mean free-surface level is located at $y=H$.}
    \label{fig:general_geometry}
\end{figure}

We represent the lock (shown in Figure \ref{fig:general_geometry}) in an idealised rectangular tank of size $L \times H_{d}  \times W$ in the $x$, $y$, and $z$ directions, respectively (with $y$ the vertical direction). The origin of the reference frame is located at the centre of the tank in the horizontal and at the bottom of the tank in the vertical. The lock gate, located at $x=0$, separates two halves of the tank such that each half has a length $L/2$ filled with water of different density up to a depth $H$. Throughout the paper constant values of $L = 6$ \unit{m}, $H_d = 0.2$ \unit{m}, $H = 0.15$ \unit{m} and $W = 0.2$ \unit{m} were used. A porous sparger is located on the tank floor ($y=0$) at the location of the lock gate ($x=0$) and spans the tank in the $z$-direction (i.e. along the width). In the $x$-direction, it has a width $d_s = 0.02$ \unit{m}, which means that it is located at $- d_s/2 \leq x \leq d_s/2$ and $-W/2 \leq z \leq W/2$. The sparger delivers an air volume flow rate per unit width $q_{air}$ to form the line bubble plume. The bubbles are assumed to have a uniform bubble diameter $d_b = 2$ \unit{mm}.

Performing a dimensional analysis for the problem yields the following dimensionless parameters,
\begin{equation}
\tilde{\text{L}} = \frac{L}{H}, \hspace{0.3cm}  \tilde{\text{W} }= \frac{W}{H},\hspace{0.3cm}
\Fr = \frac{(g \, q_{air})^{1/3}}{\sqrt{g^{\prime} H}}, \hspace{0.3cm}
\rd - 1 =\dfrac{g'}{g},\hspace{0.3cm}
\Reg = \frac{\sqrt{g^{\prime} H}H}{\nu},
\label{eq:non_dimensional_groups}
\end{equation}
where $g$ is the gravitational acceleration, $\nu$ is the kinematic viscosity of water, and $g^{\prime} = g(\rho_s - \rho_f)/\rho_f$ is the reduced gravity with $\rho_s$ and $\rho_f$ the density of salt and fresh water, respectively. Furthermore, combining $\Fr$ and $\sigma -1$ results in an additional relevant parameter: 
\begin{equation}\label{eq:q_hat}
    \q=\Fr^3(\rd -1)^{3/2}=q_{air}/(H\sqrt{g H}).
\end{equation}
This parameter has been shown to govern the entrainment by bubble plumes in water with homogeneous density \citep{fannelop1991surface}. The entrainment increases with it up to $\q\approx 1.6 \times 10^{-3}$ and remains approximately constant for larger values. 

We further introduce the effectiveness as a measure of the amount of salt water that can be blocked by the curtain compared to a case without it. It is defined as 
\begin{equation}
    E(t) = 1 - \frac{V_\textrm{bc}(t)}{V_\textrm{o}(t)},
    \label{eq:effectiveness}
\end{equation}
where $ V_\textrm{bc}(t) $ and $V_\textrm{o}(t)$ are the total volume of salt water that has infiltrated the fresh side of the lock when the bubble curtain is present and absent, respectively, and $t$ is the elapsed time after opening of the lock gate. Hence, $E$ measures the amount of saltwater that is blocked by the curtain, with $E=1$ a perfect barrier. Instead of effectiveness, \citet{keetels2011field} and \citet{ oldeman2020numerical} used the salt transmission factor $STF=V_\textrm{bc}/V_\textrm{o} = 1 - E$, but both quantities give the same information.

The volume $V_\textrm{bc}$ is obtained from mass conservation 
\begin{equation}
    V_\textrm{bc}(t) = V_\f\frac{\bar{\rho}_\f(t) - \rho_f}{\rho_s - \rho_f},
    \label{eq:v_star}
\end{equation}
with $V_\f=  L \, W \, H/2$ the volume on the freshwater side of the lock and $\bar{\rho}_\f(t)$ the average density on that same side \citep{bacot2022bubble}. For the infiltrated volume in the absence of a bubble curtain, we consider a constant flux so that
\begin{equation}
    V_\textrm{o}(t) = \dfrac{\text{C}_{\text{D}}}{3} W H \sqrt{g^{\prime}H} t,
        \label{eq:v_open}
\end{equation}
where $\text{C}_{\text{D}}\approx 0.51$ is the discharge coefficient that is obtained by fitting the flow rate obtained in simulations without a bubble curtain \citep[see][]{raaghav2025bubble}.

In general, effectiveness is a function of time, and \citet{raaghav2025bubble} showed that this dependence is very important in the curtain-driven regime, but that it becomes less important when approaching the breakthrough regime. In other words, $V_\textrm{bc}/V_\textrm{o}$ becomes a constant at earlier times. Hence, in this paper, we focus, similarly to \citet{bacot2022bubble}, only on the effectiveness at a final time $t=t_{end}$. Here, $t_{end}$ is the end time of the simulation or when the gravity current reaches the end wall of the tank. These end times are typically about three times longer than those considered by \citet{bacot2022bubble} because the tank in the simulations is three times longer. Hence, at $t=t_{end}$, $E$ has reached a constant value to a good approximation. To simplify the notation, we define $E_{end}=E(t=t_{end})$.

Determining the effectiveness $E_{end}$ of bubble curtains as a function of the parameters of the problem is a crucial question for their operation. Most previous studies extensively investigated the dependence of the effectiveness on $\Fr$ while keeping other parameters constant or ignoring their effects. For example, \citet{oldeman2020numerical}, \citet{ bacot2022bubble} and \citet{o2024effect} kept $\sigma - 1 = 0.020$, a typical value of the density difference between sea water and fresh water, while \citet{bacot2022bubble} also considered cases with $\sigma - 1 > 0.020$ even up to $\sigma - 1 \approx 0.187$
that extend beyond the limits where the Boussinesq approximation is valid. However, the density difference between the water in the lock chamber and the harbours can be much smaller ($\sigma-1\lesssim 0.010 $), since the lock chamber mostly contains brackish water \citep{weiler2026quantification,bakker2026accurately}. Then, it can be seen from the definition of $\q$ in Equation \eqref{eq:q_hat} that, for $\Fr \lesssim 1$ and $\sigma -1 \approx 0.010$, the value of $\q \leq 1 \times 10^{-3}$. Previous studies show that the entrainment depends on $\q$ \citep{kobus1968analysis, fannelop1991surface}. From the work of \citet{fannelop1991surface}, it can be inferred that, around the value of $\q \approx  1.6 \times 10^{-3}$, the entrainment by the plume become constant. In this study, we will show that this has important implications for the effectiveness of bubble curtains.

To study the dependence of $E_{end}$ on the non-dimensional parameters of the problem, we varied some of the parameters as follows. The values of $H$, $L$, and $W$ result in $\tilde{\text{L}}=40$ and $\tilde{\text{W}}=1.33$, and are kept constant for all simulations. We consider three values for $\sigma -1$ (0.008, 0.020 and 0.028), and we vary $\Fr$ between 0.3 and 1.1. This implies $2.7 \times 10^{-5} \leq \q\leq 3.7 \times 10^{-3} $. Finally, $\Reg \gtrsim 14500$, implying that all gravity currents are fully turbulent, as in ship locks. Hence, the dependence of the results on $\Reg$ is expected to be minimal. To vary the values of $\Fr$ and $\sigma -1$, we changed the salt water density $\rho_s$, while keeping the fresh water density $\rho_f = 999$ kg/$\textrm{m}^3$ constant. This value of $\rho_f$ is the reference density that corresponds to an ambient temperature of 288.5 \unit{K}. 

\section{Numerical methodology}
\label{section:numerical_methodology}
We perform Euler-Euler large-eddy simulations of a bubble curtain in a lock-exchange configuration using the approach followed by \citet{raaghav2025bubble}. In the Euler-Euler two-fluid modelling framework, the air and water phases are assumed to be interpenetrating continua, and the momentum exchange between the phases is modelled using a closure force term in the momentum equations \citep{drew1983mathematical, ishii2010thermo}. We solve the governing phase-averaged continuity equation and Navier-Stokes equations for incompressible Newtonian fluids under the Boussinesq approximation:
\begin{equation}
    \frac{\partial\alpha_{q}}{\partial t} +\boldsymbol{\nabla}\cdot (\alpha_q \vec{v}_q) = 0,
    \label{eq:bouss_continuity}
\end{equation}

\begin{equation}
    \frac{\partial(\alpha_q \vec{v}_q)}{\partial t} +\boldsymbol{\nabla}\cdot\left(\alpha_q\vec{v}_q \vec{v}_q \right) = -\frac{\alpha_q}{\rho_{0,q}} \boldsymbol{\nabla} p + \frac{\rho_q}{\rho_{0,q}} \alpha_q \vec{g} + \boldsymbol{\nabla} \cdot \left\{\alpha_q \nu_{e,q}\left[\boldsymbol{\nabla}\vec{v}_q + (\boldsymbol{\nabla}\vec{v}_q)^T - \frac{2}{3} (\boldsymbol{\nabla}\cdot\vec{v}_q)\textbf{I}\right] \right\} + \vec{F}_q.
    \label{eq:bouss_momentum}
\end{equation}
Here, the subscript $q$ denotes the phase [$g$ for gas (air) or $l$ for liquid (water)], $\alpha_q$ represents the volume fraction for each phase, $\vec{v}_q=(v_{q,x},v_{q,y},v_{q,z})$ the velocity, $\rho_q$ the density, $\rho_{0,q}$ a reference density, $p$ the shared pressure between the air and water phase, $\vec{g}= (0,-g,0)$ the gravitational acceleration, $\nu_{e,q}$ the effective kinematic viscosity (i.e. the sum of the molecular viscosity $\nu_q$ and the eddy viscosity $\nu_{t,q}$), $\textbf{I}$ the identity matrix, and $\vec{F}_q$ the inter-phase momentum transfer. In this study,  $\vec{F}_q$ includes the drag, lift, virtual, and mass closure forces. For the drag and lift force, we use the closure of \citet{schiller1933drag} and \citet{tomiyama2002transverse}, respectively. A constant coefficient equal to 0.5 was used for the virtual mass force. Since we are mainly interested in the liquid phase and to simplify the notation, we obviate the subindex $l$ for this phase so that $\nu_l=\nu$, $\rho_l=\rho$, $\alpha_l=\alpha$, $\vec{v}_l=\vec{v}$, and so on.\\

Buoyancy effects caused by salinity gradients are only considered within the liquid phase because the diffusivity of salt in the gas phase is orders of magnitude lower than in the liquid phase. We solve the phase-constrained formulation of the scalar transport equation to model the salinity transport in the liquid phase:
\begin{equation}
    \frac{\partial(\alpha s)}{\partial t} +\boldsymbol{\nabla}\cdot\left(\alpha\vec{v} s\right)-
    \boldsymbol{\nabla} \cdot \left(\alpha D_{\textit{eff}} \boldsymbol{\nabla} s\right) = 0,
    \label{eq:salt_transport}
\end{equation}
where $s$ is the intrinsic salt water fraction with $0\leq s \leq 1$ and $D_{\textit{eff}} = D + D_{t}$ is the effective diffusivity with $D$ the molecular diffusivity and $D_t$ the turbulent or eddy diffusivity. 
The turbulent diffusivity $D_t$ is related to the eddy viscosity such that
\begin{equation}
     D_t = \nu_{t}/ \text{Sc}_{\text{t}},
\end{equation}
with $\text{Sc}_{\text{t}}$ the turbulent Schmidt number. A constant value $\text{Sc}_{\text{t}}=0.7$ was used following, for example, \citet{spalding1971concentration} and  \citet{oldeman2020numerical}. For a round turbulent free jet \citet{spalding1971concentration} showed that this value yielded very good agreement between experiments and simulations in predicting the concentration fluctuation profiles. Furthermore, \citet{oldeman2020numerical} also observed very good agreement with the experiments for the current flow configuration. This justifies our choice in the present study. Note that we define $\alpha_s = \alpha s$, which is the conserved salt water fraction constrained to the liquid phase and is used for visualization of the salinity contours.

The Boussinesq approximation was used in Equation \eqref{eq:bouss_momentum} for the liquid phase to include the effects of density difference resulting from changes in salinity \citep{boussinesq1903theorie}. Hence, the variations in density are only incorporated in the gravitational term in Equation \eqref{eq:bouss_momentum}. The approximation is only valid for flows with small density variations \citep{nieuwstadt2016introduction}, which is a fair assumption in this study, since $\Delta \rho/\rho_f\leq 0.028$.

The intrinsic density of the liquid is then calculated as
\begin{equation}
    \rho = \rho_f + s(\rho_s-\rho_f) = s \rho_s + (1 - s) \rho_f,
\end{equation}
where the density of fresh water is taken as the reference density ($\rho_{0}=\rho_f$) such that the perturbation density is $\rho' = \rho - \rho_f=s(\rho_s-\rho_f)$. Variations in the kinematic viscosity of water due to differences in salt concentration are neglected.\\

The values of physical constants used in the simulations are given in Table \ref{table:physical_constants}. The values of kinematic viscosity of water $\nu$, surface tension $\gamma$ and salt diffusivity $D$ were estimated for an ambient temperature of 288.5 \unit{K}, as for the density of fresh water $\rho_f$. The air in the bubbles and on top of the water surface has a density $\rho_{g}$ and a kinematic viscosity $\nu_{g}$ estimated at the same ambient temperature. 

\begin{table}[htbp]
    \begin{center}
    \begin{tabular}{llll}
    \hline
    Physical Quantity   & Value & Units \\ 
    \hline
    $\nu$ & 1.123 $\times$ $10^{-6}$ & ($\textrm{m}^2$ $\textrm{s}^{-1}$) \\ 
    $\nu_{g}$ & 1.464 $\times$ $10^{-5}$ & ($\textrm{m}^2$ $\textrm{s}^{-1}$) \\ 
     $\rho_{g}$ & 1.223 & (kg $\textrm{m}^{-3}$) \\ 
    $\gamma$ & 0.0734 & (N $\textrm{m}^{-1}$) \\ 
    $D$ & 1.096 $\times$ $10^{-9}$ & ($\textrm{m}^2$ $\textrm{s}^{-1}$) \\ 
    $g$ & 9.81 & ($\textrm{m}$ $\textrm{s}^{-2}$)
    \\ 
    \hline
    \end{tabular}
    \caption{Values of the physical constants at temperature of 288.5 K.}
    \label{table:physical_constants}
    \end{center}
\end{table}

The governing equations \eqref{eq:bouss_continuity},  \eqref{eq:bouss_momentum}, and \eqref{eq:salt_transport} are discretized and solved numerically with the finite-volume method using OpenFOAM \citep{weller1998tensorial, jasak2007openfoam, chen2014openfoam}, where a customized version of the Euler-Euler solver module is used. Details regarding the discretization, flow initialisation and boundary conditions along with validation is given in \citet{raaghav2025bubble}. The only difference from the configuration of \citet{raaghav2025bubble} is that here,  for the temporal discretization, we use implicit Euler time integration instead of the Crank-Nicholson scheme. We made this change because we noticed that the simulations are more susceptible to numerical artifacts for low $\Fr$ values.

\section{Regimes at low $\Fr$ values - Qualitative description}
\label{section:regimes_qualitative_description}
\begin{figure}
    \centering
    \includegraphics[width=0.98\textwidth]{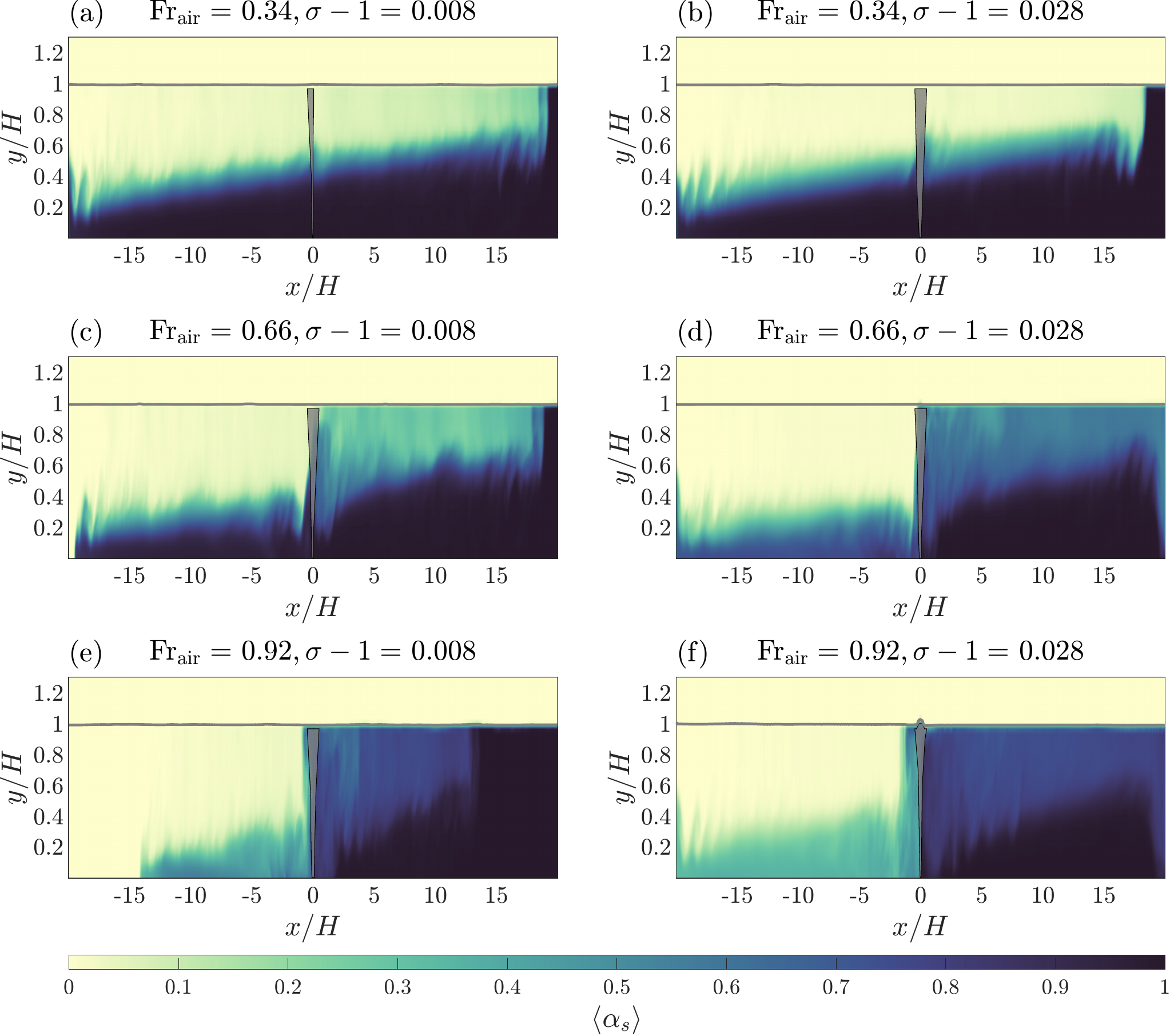}
    \caption{Width-averaged density contours represented by the salt water fraction $\langle \alpha_s \rangle$ for six simulations with different values of $\Fr$ and $\sigma$ at $t=t_{end}$. The top row shows two cases of the \emph{total breakthrough} regime, the middle row shows two cases of the \emph{diluted breakthrough} regime, and the lower row shows two cases of the \emph{curtain-driven onset} regime. The grey shaded contour depicts the location of the bubble curtain using the region with $0.5>\langle \alpha_g \rangle>5\times 10^{-4}$ and the horizontal solid grey line at $y/H \approx 1$ marks the location of the free surface.} 
    \label{fig:density_contour_regimes}
\end{figure}

\begin{figure}
    \centering
    \includegraphics[width=0.9\textwidth]{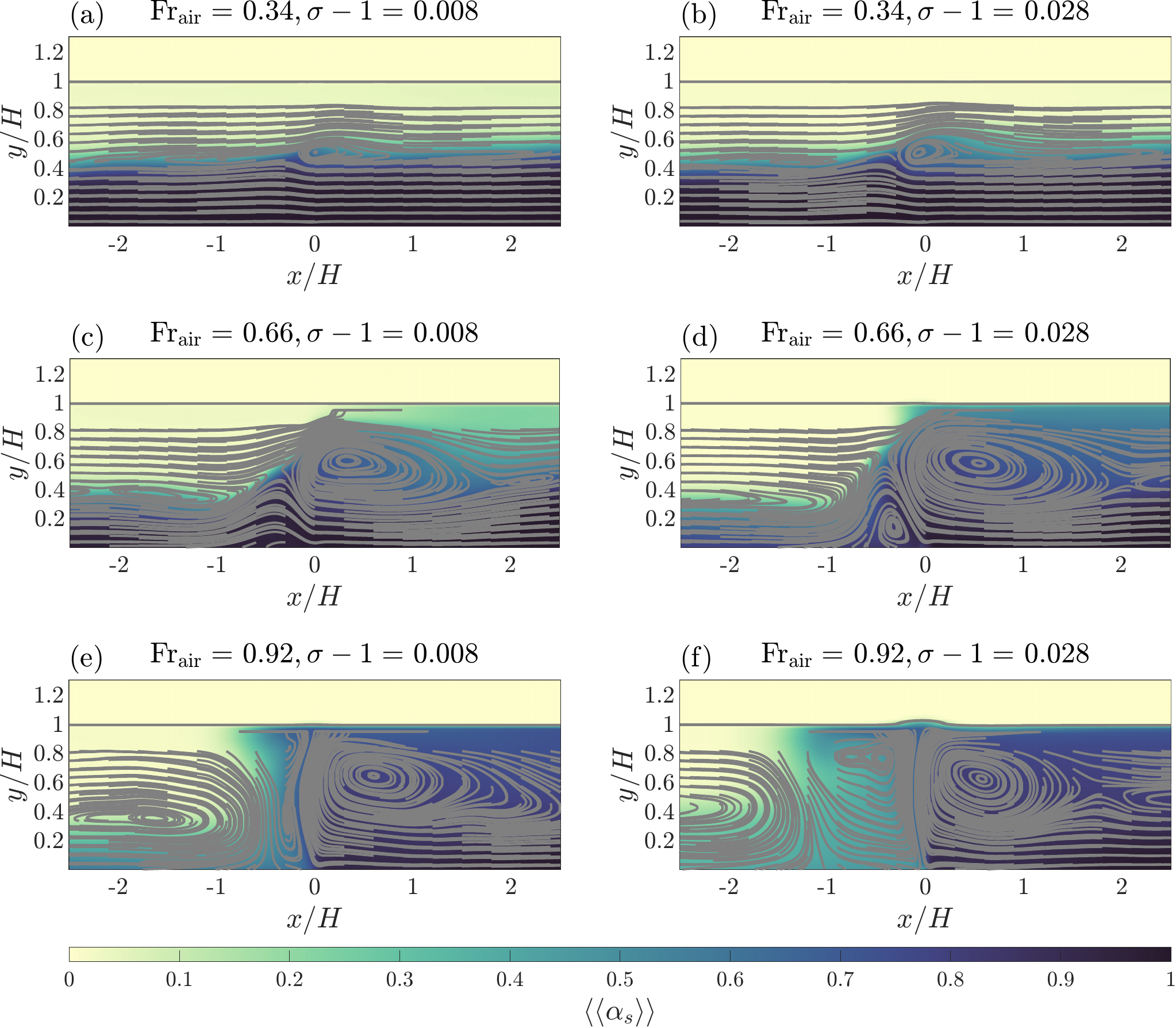}
    \caption{Width- and time-averaged density represented by the salt water fraction field $\langle \langle \alpha_s \rangle \rangle$ (colours) and streamlines (grey lines) obtained from the width- and time-averaged velocity field. The simulations are the same as in Figure \ref{fig:density_contour_regimes}, but the extent of the $x$-axis is much smaller. The horizontal solid grey line at $y/H
\approx 1$ marks the location of the free surface.}
    \label{fig:velocity_field_regimes}
\end{figure}

A first detailed observation of the density and velocity fields reveals that the breakthrough regime is actually composed of qualitatively different sub-regimes. In this section, we present these sub-regimes based on a qualitative description of the width-averaged density fields and the typical flow patterns close to the bubble curtain as shown in (Figures \ref{fig:density_contour_regimes} and \ref{fig:velocity_field_regimes}).

For the analysis of the numerical results, it is convenient to use width-average quantities, with the width-average operator is defined as 
\begin{equation}
\langle \cdots \rangle = \frac{1}{W}\int_{-W/2}^{W/2}\cdots dz.
\end{equation}
Furthermore, when studying the statistically steady state in the near-field of the plume, it is convenient to additionally average in time to remove temporal fluctuations. The width- and time-averaging operator is defined as
\begin{equation}
\langle\langle \cdots \rangle \rangle = \frac{1}{(t_{end} - t_{0})W}\int_{t_0}^{t_{end}}\int_{-W/2}^{W/2}\cdots dz dt.
\end{equation}
In all cases, we use $t_0=25$ s, which was found to be sufficient for the flow and salt concentration fields to reach a steady state in the near-field of the plume.

\subsection{Total breakthrough regime}
In the total breakthrough regime, from the density field associated with this regime (shown in Figures \ref{fig:density_contour_regimes}a and b for $\Fr$ = 0.34), we observe that the density current resembles that which would have been formed in the absence of the bubble curtain in a classical lock-exchange flow. In other words, due to the very low inertia of the bubble curtain, the density current infiltrates the curtain almost unaffected by it. Hence, the density close to the bottom is that of the salt water and the density close to the surface is that of fresh water.
 
A closer inspection of the density fields and the streamlines near the bubble curtain ($-2.5 \lesssim x/H \lesssim 2.5$) in Figure \ref{fig:velocity_field_regimes} reveals a recirculation cell around mid-depth at $x\approx0$. This cell extends in the positive $x$-direction, and for the two cases with $\Fr=0.34$ shown in Figure \ref{fig:velocity_field_regimes}, it extends vertically in the range $0.4\lesssim y/H\lesssim 0.6$. The bubble curtain entrains and lifts some dense salt water that subsequently moves in the positive $x$ direction carried by the exchange flow. The lifting of some dense fluid is reminiscent of the peeling observed in previous studies on bubble plumes in a stratified quiescent fluid \citep{socolofsky2003liquid, socolofsky2005role, yang2016large}. Due to the higher density, these fluid parcels that are lifted sink while moving in the positive $x$-direction, causing the formation of the recirculation cell. This recirculation cell induces a weak mixing of salt and fresh water in the saltwater side.

\subsection{Diluted breakthrough regime}
From the two simulations for $\Fr=0.66$ shown in Figure \ref{fig:density_contour_regimes}c and d, we observe that the dense fluid entrained by the curtain reaches the free surface. Consequently, the gravity current (exchange flow) travelling just below the surface in the positive $x$-direction is no longer composed of pure fresh water. Similarly, the salt content in the density current travelling over the bottom in the negative $x$ direction is lower than that of the salt water. Hence, we name this regime the \emph{diluted breakthrough} regime. 

The streamlines in Figure \ref{fig:velocity_field_regimes} show that the recirculation cell for $x>0$ has grown and extends over (almost) the entire water column, while still no recirculation cell is observed for $x<0$. At the free surface, the recirculation cell produces surface currents limited to the right side, contrary to the case of a classical bubble curtain in a quiescent fluid, where surface currents are observed on both sides. The lack of a surface current for $x<0$ is due to the  strength of the exchange flow near the surface in the positive $x$-direction.

Quite remarkably, in this regime, some differences are observed between the case with $\sigma-1=0.008$ and that with $\sigma-1=0.028$. On the one hand, the case with $\sigma-1=0.008$ shows less dilution and, on the other hand, the recirculation cell is slightly smaller. The differences between simulations with different values of $\sigma$ will be discussed in more detail in the following.

\subsection{Curtain-driven onset regime}
As $\Fr$ increases further, we observe further dilution in the gravity currents due to the increased mixing around the bubble curtain (see Figures \ref{fig:density_contour_regimes}e and f). Furthermore, the streamlines displayed in Figures \ref{fig:velocity_field_regimes}e and f show, in addition to a recirculation cell for $x>0$, also a small recirculation cell and surface currents for $x<0$. The simulation for $\sigma - 1=0.008$ shows a smaller recirculation cell than that for $\sigma -1=0.028$. In the curtain-driven regime, the recirculation cells on both sides of the bubble curtain are approximately equal in strength and size \citep{bacot2022bubble, raaghav2025bubble}. However, in this onset regime, even if the curtain has enough inertia to form surface currents and recirculation cells on both sides of the bubble curtain, they are strongly asymmetric. Hence, we denote this regime as the \emph{curtain-driven onset} regime.

The asymmetry between both sides is due to the fluid parcels in the surface current for $x<0$ sinking more quickly than those for $x>0$. In fact, the end of the recirculation cell for $x>0$ is not observed within the spatial range plotted in Figure \ref{fig:velocity_field_regimes}. Interestingly, \citet{abraham1973pneumatic} sketched a similar flow with a small recirculation cell for $x<0$ based on surface flow observations. However, the recirculation cell for $x>0$ was not considered, probably because its signature on the surface is not as clear. 

The gradual appearance and growth of the recirculation cell for $x<0$ with $\Fr$ shows that there is not a sharp transition between the breakthrough and the curtain-driven regime at $\Fr \approx 0.9$ \citep[as proposed by][]{bacot2022bubble}. Instead, we observe a smooth transition through the \emph{curtain-driven onset} regime.

\section{Quantitative description}
\label{section:quantitative_description}

In the previous section, we qualitatively introduced three sub-regimes: \emph{total breakthrough}, \emph{diluted breakthrough}, and \emph{curtain-driven onset}. In what follows, we provide a quantitative characterisation to more precisely capture the transitions between these sub-regimes and delineate their respective limits.

\subsection{Approach}

We aim to quantitatively describe how key characteristics of the flow and density fields change as a function of $\Fr$ and $\sigma-1$. We focus on the entrainment by the bubble curtain, the dilution by the bubble curtain,  and the emergence and growth of the recirculation cells. To quantify these changes, we define a small rectangular control volume around the bubble curtain. However, one difficulty is that the bubble curtain changes in size and shape depending on the control parameters. Hence, we must define the control volume as a function of the volume occupied by the bubble curtain in each simulation. Since there is little spanwise variation, we will consider width-averaged quantities, so that we need to define the extent of the volume in the $x$ and $y$ directions. The vertical limits, owing to pragmatic reasons, are 1.5 grid cells above the bottom at $y_-=0.05H$ and below the free surface at $y_+=0.95H$.

To define the extent of the control volume in the $x$-direction, we first determine the extent of the bubble curtain in this direction by fitting a Gaussian curve to the air volume fraction averaged in width and time $\langle\langle \alpha_g \rangle\rangle$ at every height $y$ as a function of $x$ such that 
\begin{equation}
\langle\langle\alpha_g\rangle\rangle_\textrm{fit}=A(y) \exp\left\{- \left[ \dfrac{x-B(y)}{C(y)}\right]^2\right\},
\end{equation}
with $A(y)$, $B(y)$ and $C(y)$ the fitting constants. From this fit, we define the width of the curtain as $w(y)=2\sqrt{2}C(y)$, which is equivalent to four times the standard deviation of the fitted Gaussian profile. In other words, it represents the region that encompasses 95\% of the air volume at a given height $y$. Using the width of the bubble curtain, the limits of the bubble curtain are $x=B(y)+w(y)/2$ on the right side and $x=B(y)-w(y)/2$ on the left side. The horizontal limits of the control volume for our analysis are the leftmost and rightmost points of the boundaries of the curtain:
\begin{equation}\label{eq:limit_left}
    x_-=\min[B(y)-\sqrt{2}C(y)]
\end{equation}
and
\begin{equation}\label{eq:limit_right}
    x_+=\max[B(y)+\sqrt{2}C(y)],
\end{equation}
respectively.

Although the choices we have made to define the control volume are somewhat arbitrary, we note three important characteristics that it must satisfy: 1) it must be large enough to include the entire plume, 2) its boundaries must be close enough to the plume so that structures such as the signatures of the recirculation cells are observed, and 3) it must adjust depending on the size of the curtain parameters $\sigma-1$ and $\Fr$. 
 
\begin{figure}
    \centering
    \includegraphics[width=0.98\linewidth]{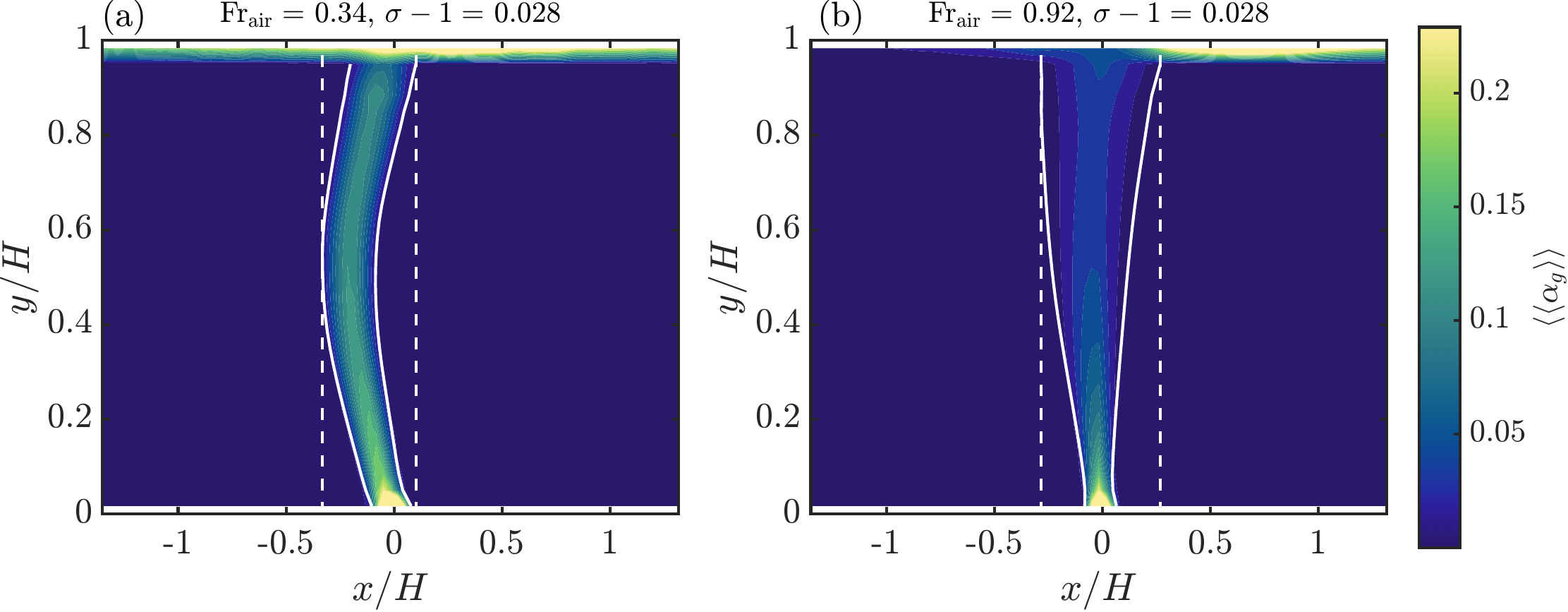}
    \caption{Two examples of the definition of the limits of the bubble curtain and the control volume used for the analysis. The colour denotes the width- and time-averaged air volume fraction $\langle\langle \alpha_g \rangle \rangle$. The solid white curves denote the limit of the bubble curtain, and the white dashed lines denote the limits $x=x_-$ and $x=x_+$ of the control volume for analysis given by Equations \eqref{eq:limit_left} and \eqref{eq:limit_right}, respectively. }
    \label{fig:examples_control_volumes}
\end{figure}

The limits of the curtain and the control volume for two examples are shown in Figure \ref{fig:examples_control_volumes}. As can be seen, the limits of the curtain (solid lines) coincide well with what would be expected from visual inspection of the width-averaged air volume fraction. The boundaries of the control volume are shown with the box with white dashed lines in Figure \ref{fig:examples_control_volumes}. For the two examples, we chose one simulation in the total breakthrough regime and one in the curtain-driven onset regime (the same simulations as shown in Figures \ref{fig:density_contour_regimes}b, \ref{fig:density_contour_regimes}f, \ref{fig:velocity_field_regimes}b and \ref{fig:velocity_field_regimes}f). We highlight two observations that stress the importance of adjusting the size of the control volume for each simulation. First, the bubble curtain in the total breakthrough regime is curved due to the strength of the gravity current. For this reason, the limit $x_-$ is not given by the bubble curtain limit close to the free surface. Second, we observe that the width of the curtain for the simulation in the curtain-driven onset regime grows faster with height. 

Due to the gravity current and the recirculation cells, it is difficult to obtain the structure of the bubble curtain from the vertical velocity. In other words, it is difficult to obtain the momentum plume. However, it has previously been observed for the homogeneous case that the spread of momentum is proportional to the spread of air in the curtain. Hence, it is possible to estimate, following \citet{ditmars1974analysis}, the entrainment coefficient of the bubble curtain $\beta$ from the width of the bubble curtain $w(y)$, such that
\begin{equation}
(y-y_v)\beta =\frac{1}{\lambda} \sqrt{\frac{\pi}{32}} w(y),
\end{equation}
where $\lambda$ is the constant that represents the ratio of the lateral spread of air (bubbles) to momentum and $y_v$ is the height of the virtual origin of the curtain. The value $\lambda=0.85$ proposed by \citet{fannelop1991surface} gives good qualitative results (a quantitative estimate of the error is difficult due to the gravity current and the recirculation cells). Notice, that $\beta$ is a constant if $w(y)$ grows linearly with $y$. To verify that this is a good approximation and obtain the value of $\beta$, we make a linear fit of $w$ as a function of $y$ of the form $w(y)=ay+b$ with $a$ and $b$ the fitting parameters. Hence, $\beta = (a/\lambda)\sqrt{\pi/32}$ and $y_v=-b/a$.  

Now that the control volume is defined, we study the dilution and the flow around the bubble curtain by extracting, at both $x=x_-$ and $x=x_+$, vertical profiles for the width and time averages of the velocity in the $x$-direction, $v_x(y)$, and the density profile, $\rho(y)$. For simplicity and to compare the simulations with different values of $\sigma -1$, we define the normalised width- and time-averaged density

\begin{equation}
\hat{\rho}=\dfrac{\langle \langle \alpha \rho\rangle \rangle - \langle \langle \alpha \rangle \rangle\rho_f}{\langle \langle \alpha \rangle \rangle\Delta \rho},
\end{equation}
so that $\hat{\rho}=0$ for fully fresh water and $\hat{\rho}=1$ for fully salt water.

In the following section, we present results on the entrainment coefficient, the dilution by the bubble curtain, and the shape of the width- and time-averaged profiles of the horizontal velocity at $x=x_-$ and $x=x_+$  as a function of the control parameters. In this way, we identify the sub-regimes and their limits. Finally, we discuss how the effectiveness varies as a function of the control parameters and across the sub-regimes.

\subsection{Sub-regime characteristics and limits}

\subsubsection{Entrainment coefficient}

\begin{figure}
    \centering
    \includegraphics[width=0.5\linewidth]{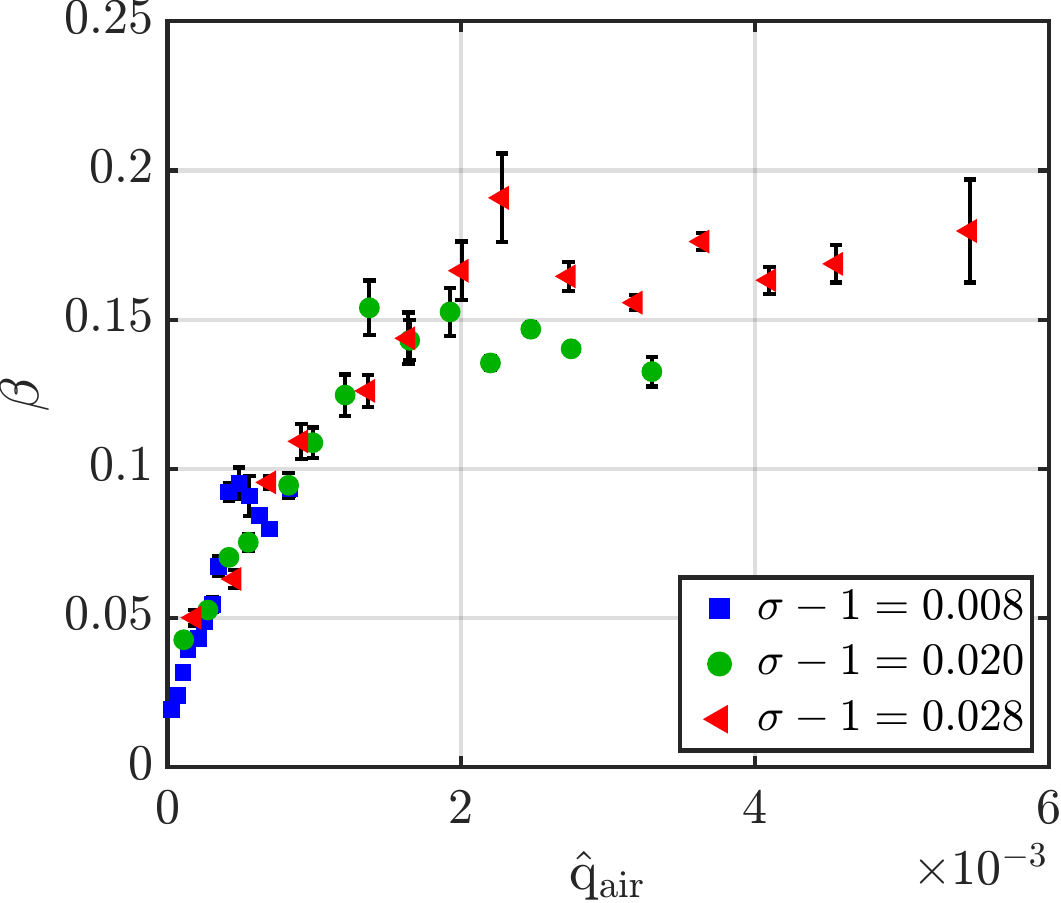}
    \caption{Entrainment coefficient $\beta$ as a function of $\q$ for all simulations. The error bars denote 95\% confidence interval}
    \label{fig:entrainment}
\end{figure}

We first discuss the dependence of the entrainment coefficient $\beta$ on the control parameters, since it impacts both the dilution and the flow around the bubble curtain. As mentioned in Section \ref{section:setup}, \citet{fannelop1991surface} found that the entrainment by the bubble curtain (a line plume) in a homogeneous fluid depends on $\q$. Figure \ref{fig:entrainment} presents the values of $\beta$ as a function of  $\q$ for all simulations. The error bars denote the 95\% confidence interval. Due to the small errors, we conclude that considering that $w(y)$ varies linearly with $y$ is a good approximation, which means that $\beta$ can be considered a constant for each simulation. However, it is important to note that $\beta$ also depends on the bubble diameter $d_b$ \citep{fraga2016influence,o2024effect}. Since $d_b$ is constant throughout this study, the values of $\beta$ reported in Figure \ref{fig:entrainment} may change for other bubble sizes and should therefore be generalised with caution.

Furthermore, to good approximation, the results for different values of $\sigma$ collapse into a single curve, implying that the $\q$ is the only parameter determining the entrainment by the bubble curtain. As expected from earlier results on round plumes \citep[e.g][]{ditmars1974analysis}, $\beta$ increases with $\q$, tending to a constant value for large enough values of $\q$. Although many previous studies computed $\beta$ for different values of $\q$ \citep{ditmars1974analysis,fannelop1991surface, beelen2024planar,lee2026comparative}, they all have been restricted to relatively small values of $\q$. The largest values of $\q$ were reached in the studies of \citet{ditmars1974analysis} and \citet{fannelop1991surface}: $\q\approx 1.12\times10^{-3}$ and $1.55\times10^{-3}$, respectively. These values are still not large enough to reach the plateau region which emerges in Figure \ref{fig:entrainment}. Hence, the present study is the first that considers such high values of $\q$ for line plumes, allowing us to estimate a critical value of $\q$ beyond which the entrainment coefficient is constant.

Our results indicate that $\beta$ tends to a constant value for $\q\gtrsim 2\times 10^{-3}$. This transition seems to agree with the general trends observed by \citet{ditmars1974analysis} and \citet{fannelop1991surface} results, even if a definitive value is difficult to determine due to the gradual change in $\beta$ and the scatter in the data. For $\q>2\times10^{-3}$, $\beta=0.160\pm0.018$, which is close to the results of \cite{ditmars1974analysis} who proposed the possibility of an asymptotic value of approximately 0.16. Additionally, these results indicate that the strength of the bubble curtain (i.e. the amount of water entrained) approximated by $K(g\, q_{air})^{1/3}H$ with $K$ a constant, as commonly used in previous studies \citep[see e.g][]{bacot2022bubble}, is only valid when $\q>2\times10^{-3}$.

For $\sigma -1 = 0.020$ and 0.028, $\q < 2 \times 10^{-3}$ for some simulations, while for $\sigma -1 = 0.008$, this is the case for all simulations. As a result, for $\Fr\approx 1$, the entrainment coefficient of the bubble curtain for $\sigma -1 = 0.008$ ($\q \approx 0.7\times10^{-3}$) is up to two times smaller than for $\sigma -1 = 0.020$ ($\q \approx 2.8\times10^{-3}$) and 0.028 ($\q \approx 4.7\times10^{-3}$).

\subsubsection{Dilution by the bubble curtain}

\begin{figure}
    \centering
    \includegraphics[width=0.98\linewidth]{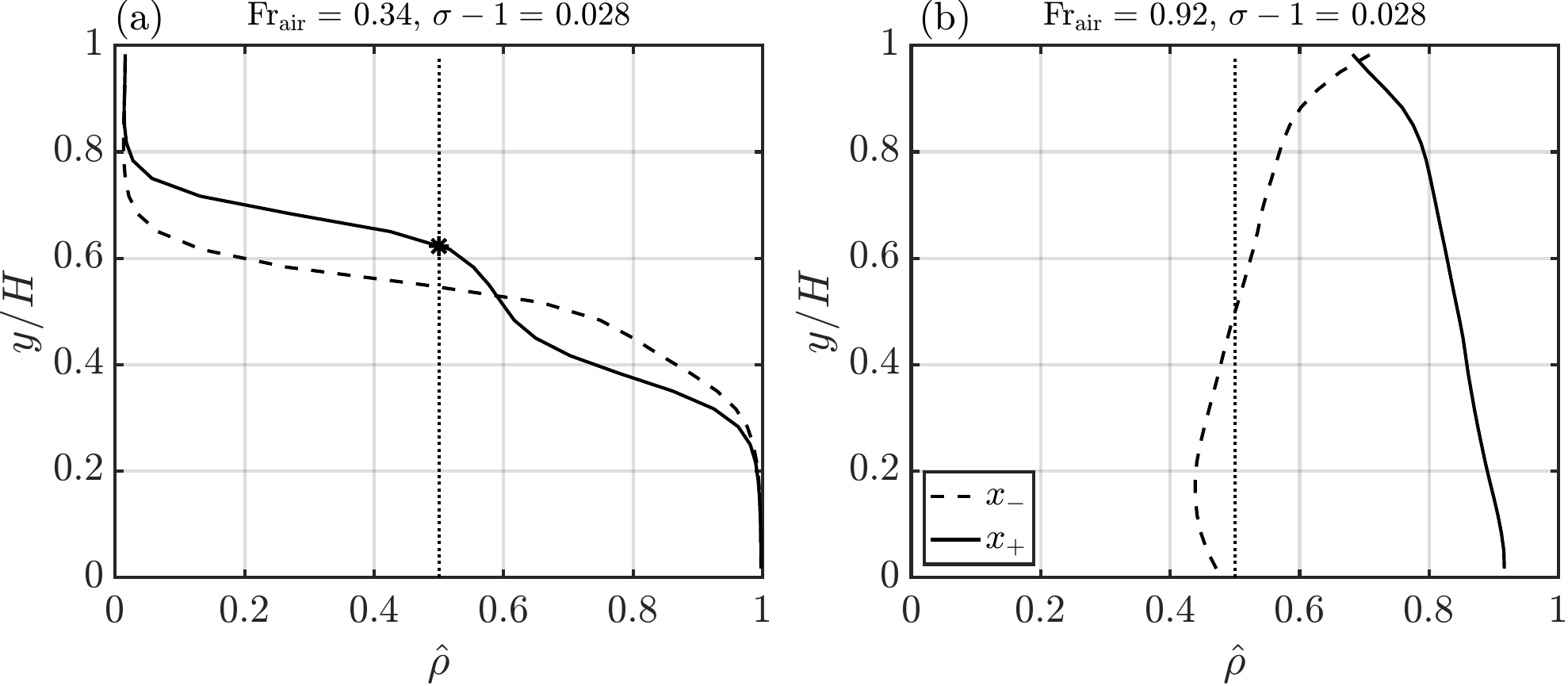}
    \caption{Normalized density profiles at $x=x_-$ (dashed line) and $x=x_+$ (solid line) for the same two simulations shown in Figure \ref{fig:examples_control_volumes}. The height $y_p$ at $x=x_+$ is marked with an asterisk. The vertical dotted line represents the value of $\hat{\rho}=0.5$.} 
    \label{fig:density_profiles}
\end{figure}

As discussed in section \ref{section:regimes_qualitative_description}, the total breakthrough regime is characterised by $\hat{\rho}=1$ close to the bottom and $\hat{\rho}=0$ close to the free surface on both sides of the bubble curtain (as seen in Figure \ref{fig:density_profiles}a). In the diluted breakthrough and curtain-driven onset regimes, the density at the bottom and close to the surface changes (with $\hat{\rho}<1$) due to the mixing induced by the bubble curtain and the recirculation cells that emerge, resulting also in more uniform density profiles (as seen in Figure \ref{fig:density_profiles}b for the curtain-driven onset regime). To distinguish the role of varying $\sigma-1$ and $\Fr$ in the density distribution around the curtain, it is useful to characterise the changes in the density distribution using the density at the corners of the control volume, i.e., at $(x_-,y_-)$, $(x_-,y_+)$, $(x_+,y_-)$, $(x_+,y_+)$. 

In addition, to characterise the ability of the bubble curtain to entrain the dense fluid upward, we compute the height $y_p$ at which the water has an intermediate density $\hat{\rho}=0.5$ at $x=x_+$. For the example in the total breakthrough regime (shown in Figure \ref{fig:density_profiles}a), mixing by the small recirculation cell on the saltwater side is reflected by a weaker density gradient around mid-height and an increase in $y_p$ to $y_p/H\approx 0.62$ (compared to $y_p/H\approx0.5$ when the bubble curtain is absent, as also observed for $x = x_-$). The height $y_p$ can be seen as a similar concept to the \emph{peel height} \citep{socolofsky2003liquid, socolofsky2005role, yang2016large}. However, the peel height of a plume in a stably stratified fluid is traditionally defined as the height at which the dye or tracer concentration is 5\% of the maximum concentration. In other words, it is the maximum height reached by the entrained fluid, reaching a neutral buoyancy level.

\begin{figure}
    \centering
    \includegraphics[width=0.98\linewidth]{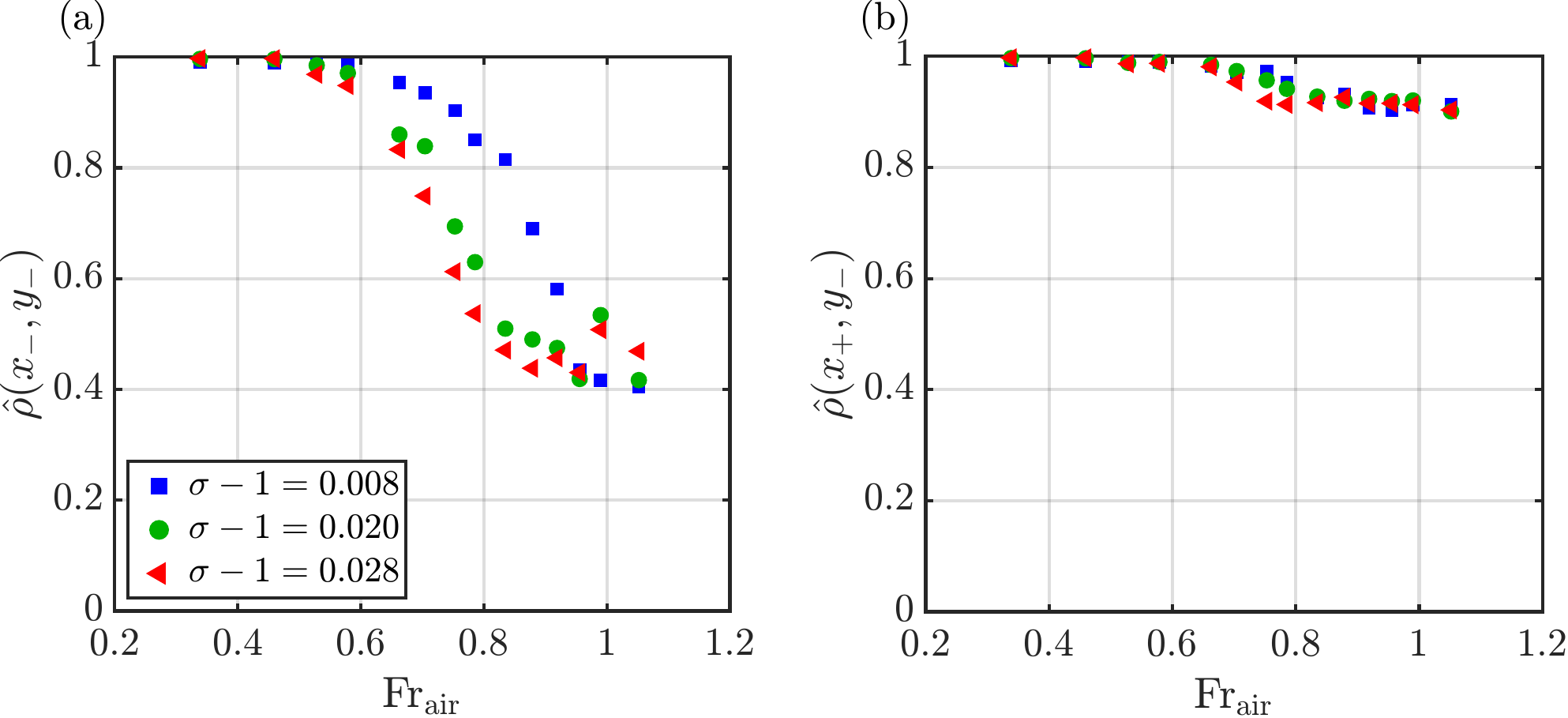}
    \caption{Normalized density at the lower corners of the control volume as a function of $\Fr$ for all simulations on a) the freshwater side $(x_-,y_-)$ and b) the saltwater side $(x_+,y_-)$.}
    \label{fig:density_characterisation_bottom}
\end{figure}

We now focus on the lower corners of the control volume: $(x_-,y_-)$ and $(x_+,y_-)$. Figure \ref{fig:density_characterisation_bottom} shows the normalised width- and time-averaged density $\hat{\rho}$ in these corners. In the lower left corner $(x_-,y_-)$ (see Figure \ref{fig:density_characterisation_bottom}a), the density decreases from $\hat{\rho}=1$ to $\hat{\rho}=0.4$ with the decrease starting at $\Fr\approx 0.6$ for $\sigma -1=0.020$ and 0.028, and at $\Fr\approx 0.7$ for $\sigma-1 = 0.008$. The start of this decrease marks the start of the diluted breakthrough regime, since it signifies that the gravity current emerging from the plume region is diluted. The fact that $\hat{\rho}$ reaches a constant value for $\Fr\gtrsim0.9$ means that there is a maximum in the dilution. 

\begin{figure}
    \centering
    \includegraphics[width=0.98\linewidth]{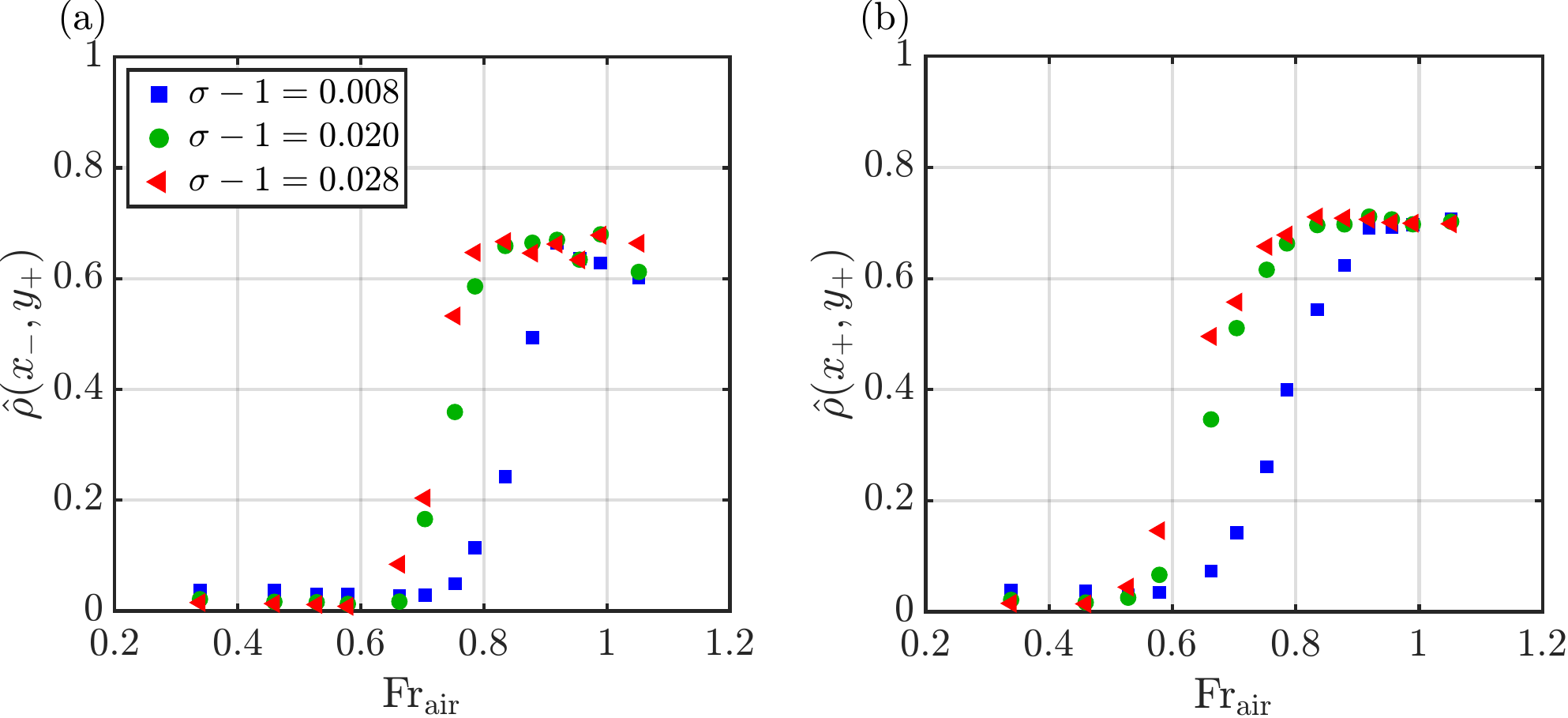}
    \caption{Normalized density at the upper corners of the control volume as a function of $\Fr$ for all simulations on a) the freshwater side  $(x_-,y_+)$ and b) the saltwater side $(x_+,y_+)$.}
    \label{fig:density_characterisation_top}
\end{figure}

Although the density and transition in the lower left corner $(x_-,y_-)$ depend on $\sigma -1$, in the lower right corner $(x_+,y_-)$ (see Figure \ref{fig:density_characterisation_bottom}b), the density is independent of $\sigma-1$. In other words, the normalised density of the salt water that reaches the bubble curtain close to the bottom on the saltwater side is independent of $\sigma-1$. However, this parameter is relevant for how this salt water is redistributed by the bubble curtain, unless $\sigma-1$ is large enough ($\sigma -1 \gtrsim 0.020$). This is a first indication that the difference in entrainment for simulations with the same value of $\Fr$ but different values of $\sigma-1$ and $\q$ affects the mixing and distribution of salt water by the bubble curtain.

The transition from the diluted breakthrough regime to the curtain-driven onset regime is marked by the formation of the recirculation cell on the freshwater side, so we focus now on the density in the two upper corners of the control volume (see Figure \ref{fig:density_characterisation_top}). Here, the normalised density increases from $\hat{\rho}\approx 0$ to $\hat{\rho}\approx 0.7$. In the upper right corner $(x_+,y_+)$, where a recirculation cell is always present, this increase occurs at similar values of $\Fr$ ($\Fr\approx 0.7$ for $\sigma-1=0.008$, and $\Fr\approx 0.6$ for $\sigma-1=0.020$ and 0.028) as the decrease in density in the lower left corner (see Figure \ref{fig:density_characterisation_bottom}a). However, the increase in the upper left corner occurs for slightly larger values of $\Fr$ ($\Fr\approx 0.8$ for $\sigma-1=0.008$, and $\Fr\approx 0.7$ for $\sigma-1=0.020$ and 0.028); see Figure \ref{fig:density_characterisation_top}a.

\begin{figure}
    \centering
    \includegraphics[width=0.49\linewidth]{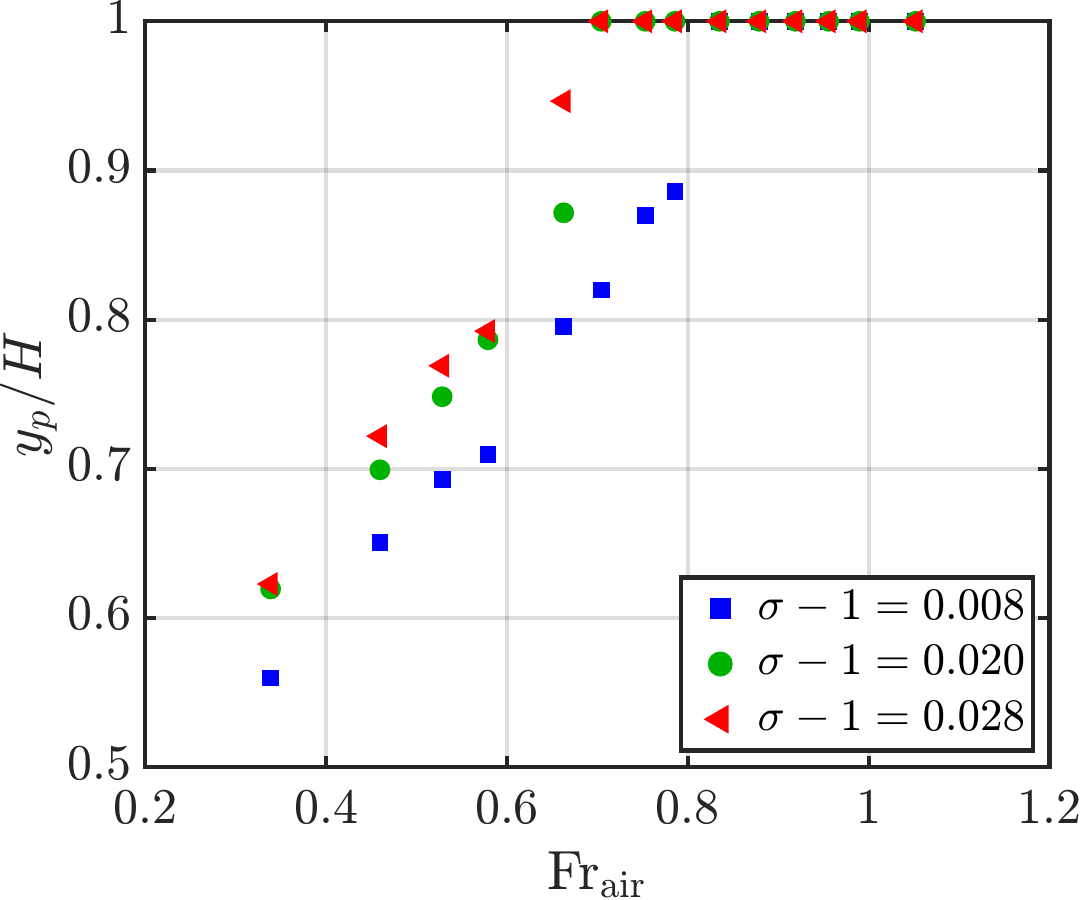}
    \caption{Normalized peel height $y_p/H$ as a function of $\Fr$ for all simulations.}
    \label{fig:peel_height}
\end{figure}

The start of the density increase on the upper left corner signals the formation of the recirculation cell on the fresh water side, and hence the transition to the curtain-driven onset regime. The formation of this recirculation cell is also observed in the value of $y_p$ (the height reached by water with $\hat{\rho}=0.5$ at $x=x_+$) as shown in Figure \ref{fig:peel_height}. In general, $y_p/H$ increases almost linearly with $\Fr$ until suddenly (at $\Fr\approx 0.8$ for $\sigma-1=0.008$, and $\Fr\approx 0.7$ for $\sigma-1=0.020$ and 0.028), its value increases to $y_p/H=1$. Again, we see a difference between the results for $\sigma-1\geq0.020$ and $\sigma-1=0.008$, where the increase of $y_p$ is slower for $\sigma-1=0.008$, again pointing to the effect of the reduced entrainment by the bubble curtain for simulations with this value of $\sigma-1$. 

In summary, the density at the four corners of the control volume and the height $y_p$ suggest that the flow transitions from the total breakthrough regime to the diluted breakthrough regime at $\Fr\approx0.6$ for $\sigma-1\gtrsim0.020$ and at $\Fr\approx0.7$ for $\sigma-1=0.008$. Furthermore, the transition between the diluted breakthrough regime and the curtain-driven onset regime occurs at $\Fr\approx0.7$ for $\sigma-1\gtrsim0.020$ and at $\Fr\approx0.8$ for $\sigma-1=0.008$. These limits point to a very small region of the parameter space for the diluted breakthrough regime. Furthermore, although the normalised density at the corners of the control volume is approximately  constant within the total breakthrough and curtain-driven onset regimes, other important flow characteristics vary within these regimes. We have already seen how $y_p/H$ increases linearly within the total breakthrough regime. In the following section, we will study how the velocity varies within these regimes.

\subsubsection{Velocity profiles}

\begin{figure}
    \centering
    \includegraphics[width=0.98\linewidth]{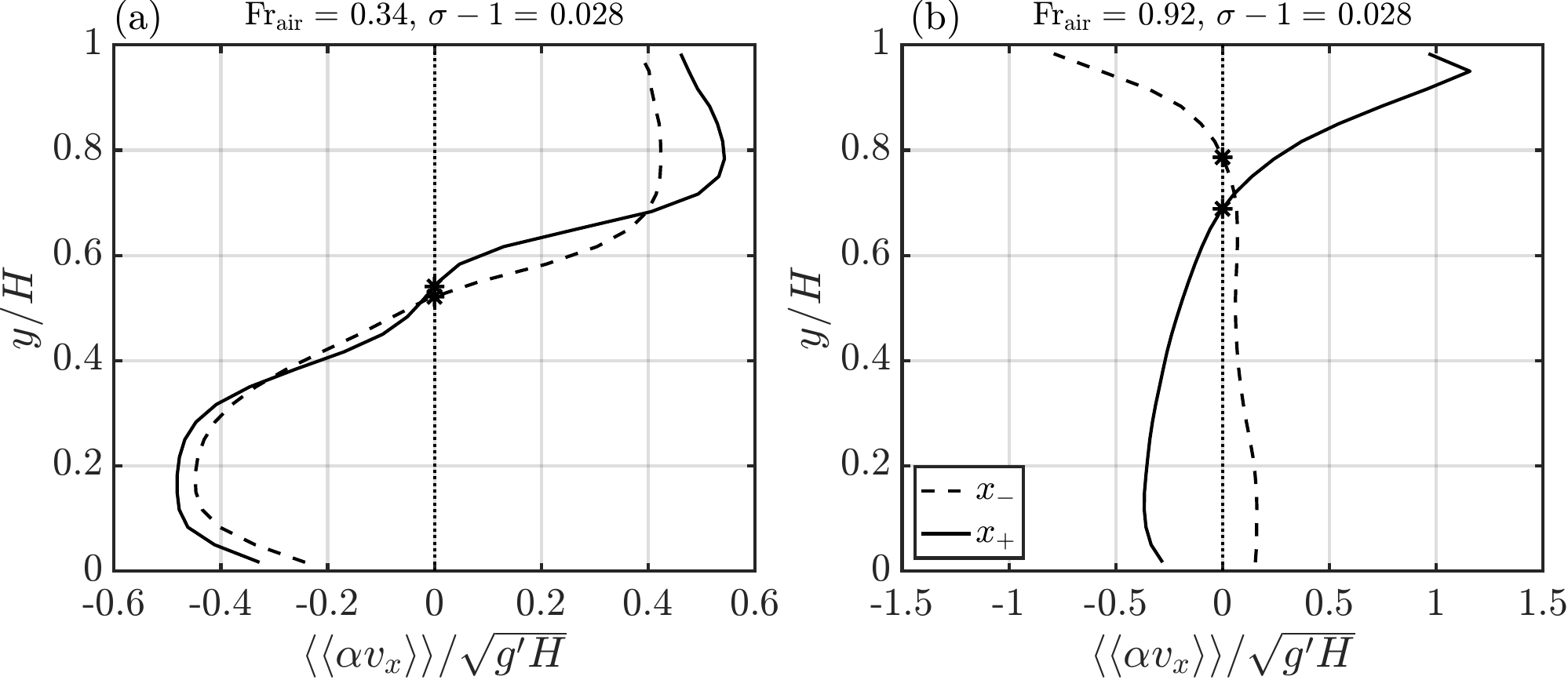}
    \caption{Profiles of the dimensionless width- and time-averaged velocity component $\langle\langle \alpha v_x \rangle\rangle/\sqrt{g^{\prime}H}$ at $x=x_-$ (dashed line) and $x=x_+$ (solid line) for the same two simulations shown in Figure \ref{fig:examples_control_volumes}. The vertical dotted line represents $\langle \langle \alpha v_x \rangle \rangle=0$ and the uppermost location where $\langle \langle \alpha v_x \rangle \rangle=0$ for each profile is marked with an asterisk.}
    \label{fig:velocity_profiles}
\end{figure}

For the velocity, we expect from the qualitative descriptions of the regimes in section \ref{section:regimes_qualitative_description} a change from an almost undisturbed lock exchange flow in the total breakthrough regime to a flow with recirculation cells on both sides of the curtain in the curtain-driven onset regime. This change can be seen in the velocity profiles at $x=x_+$ and $x=x_-$ in the total breakthrough and the curtain-driven onset regimes in Figure \ref{fig:velocity_profiles}. For the profiles in the total breakthrough regime (see Figure \ref{fig:velocity_profiles}a), the flow close to the bottom remains in the negative $x$-direction and close to the surface in the positive $x$-direction, which is the characteristic flow structure of the exchange flow. In the curtain-driven onset regime (shown in Figure \ref{fig:velocity_profiles}b), the flow directions are reversed for the profile at $x=x_-$. 

\begin{figure}
    \centering
    \includegraphics[width=0.98\linewidth]{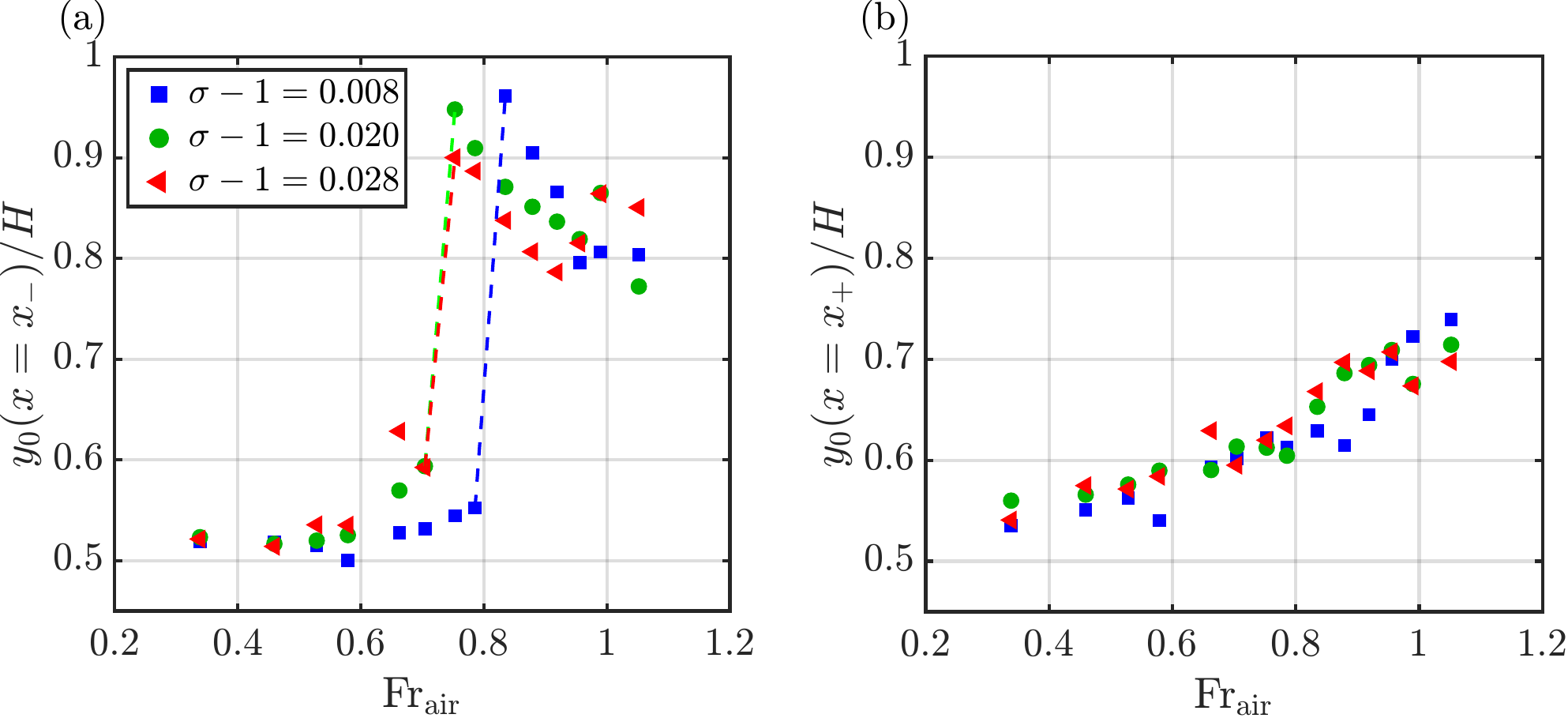}
    \caption{Height $y_0$ at which the width- and time-averaged velocity $\langle \langle v_x \rangle \rangle$ changes sign for each side of the control volume on a) freshwater side $(x=x_-)$ and b)  saltwater side $(x=x_+)$. The dashed lines are a visual guide marking the sharp increase in $y_0$ when the recirculation cell emerges on the freshwater side.}
    \label{fig:y0}
\end{figure}

Furthermore, the height at which the velocity changes sign (denoted as $y_0$ and marked with asterisks in Figure \ref{fig:velocity_profiles}) increases from $y_0\approx 0.5 $, as in a lock exchange flow \citep[see e.g.][]{shin2004gravity}, to $y_0\approx 0.75$ in the curtain-driven onset regime like for a bubble curtain in a homogeneous fluid \citep{bulson1961currents}. Hence, to characterise the changes in the velocity profile, we focus on the variation of $y_0$ as a function of $\Fr$ as shown in Figure \ref{fig:y0}.

The height at which the velocity changes sign for both $x=x_-$ and $x=x_+$ as a function of $\Fr$ is shown in (Figures \ref{fig:y0}a and b), respectively. We observe that, indeed, $y_0(x=x_-)\approx 0.5$ for $\Fr\lesssim0.5$. For higher values of $\Fr$, first, $y_0(x=x_-)$ gradually increases, and then there is a sharp increase to a value close to unity.  This jump is associated with the emergence of the recirculation cell on the freshwater side in the upper part of the water column. The values of $\Fr$ at which we observe the changes in $y_0$ coincide with the transition values observed in the density as discussed previously with the simulations with $\sigma-1\geq 0.020$ showing similar results, while the transitions for $\sigma -1=0.008$ occur at slightly higher $\Fr$ values. Within the curtain-driven onset regime, the recirculation cell grows with $\Fr$ as can be seen from decreasing values of  $y_0(x=x_-)$ for $\Fr\gtrsim 0.75$ for $\sigma -1\geq 0.0020$ and for $\Fr\gtrsim0.8$ for $\sigma-1=0.008$.

On the saltwater side, $y_0(x=x_+)$ increases almost linearly from  $y_0(x=x_+)\approx 0.5$ for $\Fr\approx 0.3$ to $y_0(x=x_+)\approx 0.7$ for $\Fr\approx 1$; see Figure \ref{fig:y0}b. On this side, the direction of the exchange flow close to the surface is the same as for the surface current driven by the curtain. Hence, the flow induced by the plume and the gravity current strengthen each other. This complementarity results in a gradual increase in size of the recirculation cell and in the value of $y_0$ instead of a sharp increase like in the freshwater side (see Figure \ref{fig:y0}a). Furthermore, we note that no difference is observed in $y_0(x=x_+)$ for different values of $\sigma-1$, suggesting a compensation mechanism for the lower entrainment.

\subsubsection{Effectiveness}

The previous results imply that the flow and the density distribution around the bubble curtain depend on both $\Fr$ and $\sigma-1$ (or $\q$), and that it is not possible to find one single governing parameter. Now, the key question for the applications of bubble curtain in ship locks is if and how the effectiveness is also affected by the changes in the different governing parameters. 

\begin{figure}
    \centering
    \includegraphics[width=0.5\linewidth]{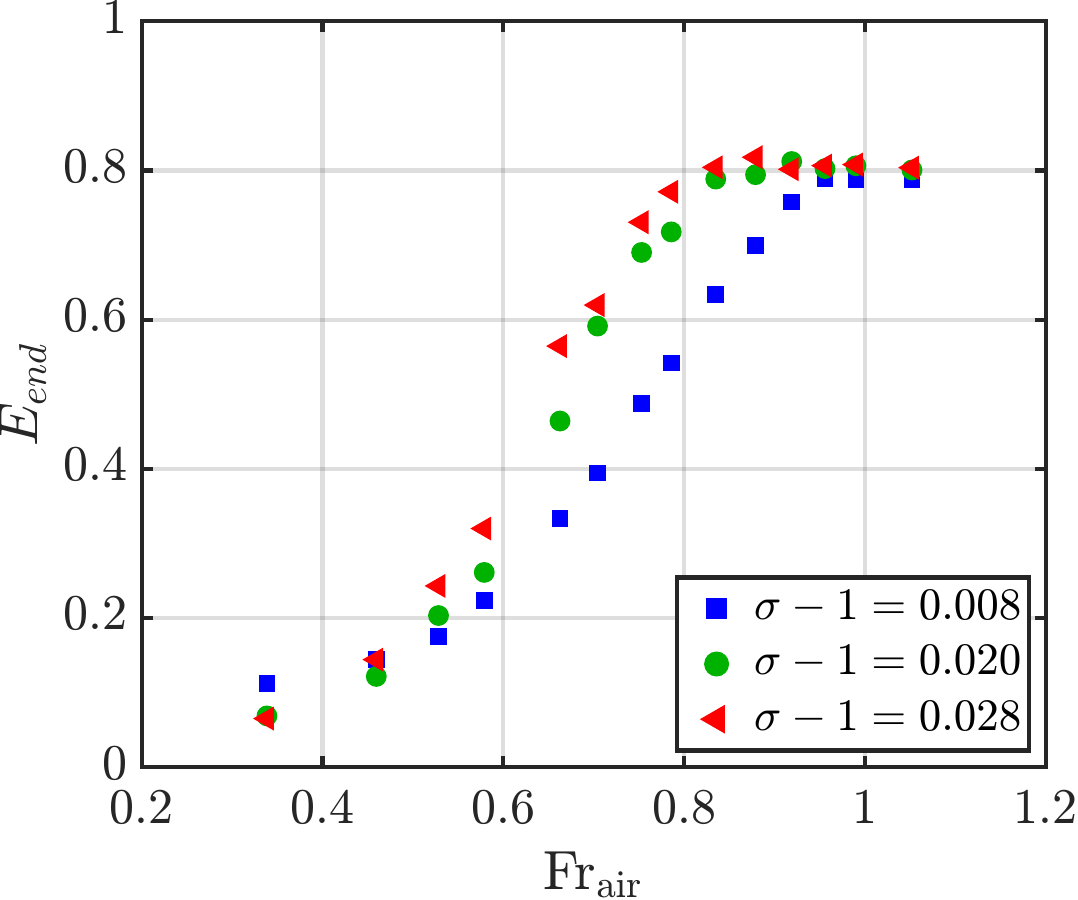}
    \caption{Effectiveness as a function of $\Fr$ for all simulations.}
    \label{fig:effectiveness}
\end{figure}

Figure \ref{fig:effectiveness} shows the effectiveness as a function of $\Fr$ for all simulations. First, for $\Fr\lesssim0.5$, the effectiveness seems to vary little with $\Fr$. Later, for $0.5\lesssim\Fr\lesssim 0.8$, the effectiveness increases  (almost linearly) until the maximum effectiveness $E\approx0.8$ is reached. In this range of $\Fr$ values, the results of the simulations with $\sigma-1=0.020$ and 0.028 show minor differences, while the simulations with $\sigma-1=0.008$ have a markedly lower effectiveness than those for $\sigma-1\gtrsim0.020$. Furthermore, the maximum value $E\approx0.8$ is reached at $\Fr\approx 0.8$ for $\sigma-1=0.020$ and 0.028, while it is reached at $\Fr\approx 1$ for $\sigma-1=0.008$. The fact that changes in the trend of $E$ occur at higher values of $\Fr$ for $\sigma-1=0.008$ compared to $\sigma-1\geq0.020$ is consistent with changes in the density and velocity profiles in the limits of the control volume as discussed previously. In general, we observe that for $\Fr\gtrsim 0.8$, the effectiveness depends only on $\Fr$ for large values of $\sigma-1$ and the effectiveness values collapse onto a single curve, while for lower values (e.g. $\sigma-1=0.008$), the effectiveness is lower. This can be attributed to a lower entrainment by the bubble curtain as discussed earlier and shown in Figure \ref{fig:entrainment}.

\section{Discussion}
\label{section:discussion}

For practical applications, it is desired to operate bubble curtains close to the optimum or at slightly lower $\Fr$ values to require less energy without a major loss in effectiveness. Hence, it is important to understand how the effectiveness and its optimal value depend on the parameters of the problem. From the early work of \citet{abraham1962reduction} to the studies of \citet{oldeman2020numerical}, \citet{bacot2022bubble}, and \cite{o2024effect}, the effectiveness of bubble curtains for reducing salt intrusion has been considered exclusively as a function of $\Fr$. However, our results show that the effectiveness is also a function of an additional parameter (either $\sigma-1$ or $\q$). This is of importance for small values of $\Fr$ (i.e. in the breakthrough regime) because, in this range, low values of $\sigma$ translate into low values of $\q$, implying that the entrainment coefficient of the bubble curtain is not constant but highly dependent on $\q$. 

The value $\sigma-1=0.020$ is typical for the density difference between sea water and fresh water, and similar values have been used in recent experimental and numerical studies \citep{oldeman2020numerical,bacot2022bubble,o2024effect, raaghav2025bubble}. For this and larger values of $\sigma-1$, the effectiveness of the bubble curtain tend to collapse into a single curve because the entrainment by the bubble curtain is independent of $\q$. However, in ship locks, $\sigma-1$ tends to be a factor of two smaller ($\sigma-1\approx 0.010$) \citep{weiler2026quantification,bakker2026accurately}. For these smaller values, there is continuous variation in effectiveness as the entrainment of the bubble curtain varies with $\q$ (i.e. it varies with $\sigma-1$ for a constant value of $\Fr$). Importantly, for $\sigma-1\approx10$, the effectiveness in the breakthrough regime can be up to 20\% smaller and the maximum effectiveness occurs for higher values of $\Fr$ compared to situations where $\sigma-1 \gtrsim 0.020$. These results suggest that bubble curtains in locks further inland where $\sigma-1$ might have much lower values would be less effective and optimum effectiveness would occur at higher $\Fr$ values.

In general, the dependence of the entrainment coefficient on $\q$ in the curtain-driven regime is of secondary importance. \citet{raaghav2025bubble} completely neglected this dependence by considering only simulations with $\q>4\times10^{-3}$. Even for the case with $\sigma-1=0.010$, the entrainment coefficient becomes constant (i.e. $\q>2\times10^{-3}$) for $\Fr\approx1.25$, barely into the range of $\Fr$ values considered by \citet{raaghav2025bubble}. However, it is important to consider this dependence when interested in optimal effectiveness ($\Fr\approx 1$) or for the intermediate $\Fr$ range ($0.5 \leq \Fr \leq 0.8$) where the regime characteristics and the effectiveness depend on $\q$. 

Furthermore, our results show that the flow in the breakthrough regime displays important qualitative differences for different values of the control parameters. We describe three qualitatively different regimes dependent on two control parameters: $\Fr$ and either $\sigma-1$ or $\q$. Transitions between different regimes occurred at different $\Fr$ values because the entrainment by the bubble plume depends on $\q$. Previous work proposing theoretical estimates for the effectiveness as a function of $\Fr$ \citep[see e.g.][]{abraham1973pneumatic,bacot2022bubble} has not considered the dependence of the entrainment coefficient on $\q$. For example, \citet{bacot2022bubble} used the constant value proposed by \cite{paillat2014entrainment}. \citet{abraham1973pneumatic} used a range of values for the entrainment coefficient but considered it independent of $\Fr$. In addition, to build the theoretical model, \citet{abraham1973pneumatic} used the flow characteristics based on the results of \cite{bulson1961currents} in the case of homogeneous water. However, our results show that the flow structure in the breakthrough regime (e.g., the size of the recirculation cells and the height of the surface currents) varies with the parameters of the problem. Hence, we expect that the detailed new knowledge provided by our simulations can be used to develop new analytical models for the effectiveness of bubble curtains in the breakthrough regime. 

We remark that the water column depth used in this study ($H=0.15$ \unit{m}) is typical for lab-scale configurations where a constant bubble size approximation is valid. For field-scale ship locks with a much larger depth, $d_b$ can vary due to compressibility effects as they rise or due to coalescence and breakup that occurs at high gas flow rates. This can be especially relevant because the entrainment characteristics of the plume, which affects the effectiveness of the curtain, depend on the bubble size \citep{o2024effect}. With changes in $d_b$, the slip velocity $u_s$ of the bubble also varies. In that case, the dimensionless slip velocity $u_s/(gq_{air})^{1/3}$ is an additional non-dimensional parameter \citep{dissanayake2021bubble}. For the current study, since $d_b$ is kept constant throughout, $u_s$ is also a constant. In this way, the present work can serve as a baseline to understand and quantify the role of varying bubble size on the effectiveness of bubble curtains. Additional simulations varying $d_b$ (or  $u_s$) would be needed to disentangle and to isolate the relative effects of  $u_s$ and $(gq_{air})^{1/3}$ which is beyond the scope of the present study.

\section{Conclusions}
\label{section:conclusions}
We performed large-eddy simulations of a bubble curtain in a lock-exchange configuration for $\Fr \lesssim 1$. This range covers the breakthrough regime, extending into the onset of the curtain-driven regime. We report new insights into the flow characteristics and show that what has been deemed the ``breakthrough regime" by \citet{bacot2022bubble} is composed of qualitatively different sub-regimes. We refer to these as the \emph{total breakthrough}, the \emph{diluted breakthrough}, and the \emph{curtain-driven onset} regimes (in order of increasing $\Fr$ values). 

In the total breakthrough regime, the exchange flow is nearly unaffected by the presence of the bubble curtain. In the diluted breakthrough regime, the bubble curtain entrains the salt water from the bottom, and a large recirculation cell is formed on the saltwater side of the bubble screen. Hence,  salt and fresh water are mixed, and the gravity current emanating from the bubble curtain on the freshwater side of it has an intermediate density. Finally, in the curtain-driven onset regime, an additional recirculation cell appears on the freshwater side of the curtain, and salt water is entrained by the curtain all the way to the surface.

Importantly, the transitions between these regimes do not depend only on $\Fr$, but also on an additional non-dimensional parameter: either $\sigma-1$ or $\q$ (only two of the three parameters are independent). In the present work, we thoroughly examined the entrainment characteristics of the curtain and its dependence on $\q$. To our knowledge, this is the first study showing the asymptotic behaviour of the entrainment coefficient for line bubble plumes as a function of $\q$ and allowing to estimate the critical value of $\q \approx 2 \times 10^{-3}$ beyond which the entrainment coefficient becomes a constant. The fact that the entrainment coefficient depends on $\q$, when $\q\lesssim 2 \times 10^{-3}$, means that the effectiveness of bubble curtains is controlled by both $\Fr$ and $\q$.

The previous conclusions have important consequences for the common approach used to understand the effectiveness of bubble curtains, where $\Fr$ is said to characterise the relative strength of the bubble curtain and the gravity current. In such an approach, the strength of the bubble curtain is given by the amount of water entrained, approximated as $K(g\, q_{air})^{1/3}H$ with $K$ a proportionality constant. Our results imply that this approximation is only valid when $\q$ is larger than a critical value ($\q \gtrsim 2 \times 10^{-3}$). This is commonly not the case for small values of the density difference occurring in real systems ($\sigma-1\approx0.010$). Other effects such as expansion and changes in slip velocity might play an additional role \citep[see, e.g.,][]{Wuest1992,dissanayake2021bubble}. The limited validity of the constant entrainment coefficient approximation should be considered in the application of bubble curtains to mitigate saltwater intrusion.

\section*{Acknowledgements}
The authors thank both Dr. Rob Uittenbogaard and Dr. Tom O'Mahoney for the greatly insightful and stimulating discussions on various aspects of the topic of mitigating saltwater intrusion in locks using bubble curtains.

\section*{Funding}
The authors thank the Dutch Research Council (NWO) for financial support
through the NWO / TTW Perspectief Program `SaltiSolutions' P18-32, in particular Project 2 `Data and CFD for solutions' (2022/TTW/01344701). The simulations were run on the Dutch National
Supercomputing Facility SURF through an NWO Domain Science grant (2024.006).

\section*{Notation}
$A  =$ fitting constant of the Gaussian curve representing the amplitude (--)

$B  =$ fitting constant of the Gaussian curve representing the location of the peak  (m)

$C  =$ fitting constant (width parameter) of the Gaussian curve (m)

$d_b  =$ bubble diameter (m)

$d_s  =$ sparger width (m)

$D  =$ molecular diffusivity ($\textrm{m}^{2}$ $\textrm{s}^{-1}$)

$D_t  =$ turbulent diffusivity ($\textrm{m}^{2}$ $\textrm{s}^{-1}$)

$D_{eff}  =$ effective diffusivity ($\textrm{m}^{2}$ $\textrm{s}^{-1}$)

$\text{C}_{\text{D}}  =$ discharge coefficient (--)

$E  =$ Effectiveness of the bubble curtain (--)

$\Fr = $ Froude number of the bubble curtain (--)

$F_{q}  =$ total interphase momentum exchange term on phase $q$ ($\textrm{m}$ $\textrm{s}^{-2}$)

$g  =$ gravitational acceleration (m $\textrm{s}^{-2}$)

$g^{\prime}  =$ reduced gravity (m $\textrm{s}^{-2}$)

$H_d  =$ height of the tank/lock (m)

$H  =$ height of water column in the lock (m)

$I  =$ identity matrix (--)

$K  =$ proportionality constant (--)

$L  =$ length of the lock (m)

$\tilde{\text{L}}  =$ dimensionless length of the lock (--)

$p = $ shared pressure between air and water phase (Pa)

$q_{air}  =$ air flow rate per unit width ($\textrm{m}^{2}$ $\textrm{s}^{-1}$)

$\q  =$ dimensionless air flow rate per unit width (--)

$\Reg  =$ Reynolds number of the gravity current (--)

$s  =$ intrinsic salt water fraction (--)

$\text{Sc}_{\text{t}}  =$ turbulent Schmidt number (--)

$STF  =$ Salt transmission factor (--)

$t =$ elapsed time after opening of the lock gate (s)

$t_{end} =$ time when the gravity current reaches the end wall of the lock (s)

$v_{q}  =$ intrinsic velocity of phase $q$ ($\textrm{m}$ $\textrm{s}^{-1}$)

$V_{\textrm{bc}}  =$ total volume of salt water infiltrated the fresh side of the lock when the bubble curtain is present ($\textrm{m}^{3}$)

$V_{\textrm{o}}  =$ total volume of salt water infiltrated the fresh side of the lock when the bubble curtain is absent ($\textrm{m}^{3}$)

$V_\f  =$ volume on the freshwater side of the lock ($\textrm{m}^{3}$)

$w  =$ width of the bubble curtain (m)

$W  =$ width of the lock (m)

$\tilde{\text{W}}  =$ dimensionless width of the lock (--)

$x_-  =$ leftmost point of the curtain boundary (m)

$x_+  =$ rightmost points of the curtain boundary (m)

$y_-  =$ location of first grid point above the bottom wall (m)

$y_+  =$ location of first grid point below the free surface (m) 

$y_p  =$ peel height equivalent (m)

$y_v  =$ height of the virtual origin of the curtain (m)

$\alpha = \alpha_l =$ volume fraction of water (-)

$\alpha_g =$ volume fraction of air (-)

$\alpha_s =$ phase-constrained salt water fraction (-)

$\beta =$ entrainment coefficient (--)

$\lambda =$ ratio of the lateral spread of air to momentum (--)

$\nu = \nu_l =$ kinematic viscosity of water ($\textrm{m}^{2}$ $\textrm{s}^{-1}$)

$\nu_g =$  kinematic viscosity of air ($\textrm{m}^{2}$ $\textrm{s}^{-1}$)

$\rho =$ intrinsic liquid density (kg $\textrm{m}^{-3}$)

$\hat{\rho} =$ normalised width- and time-averaged density (--)

$\rho_g =$ air density (kg $\textrm{m}^{-3}$)

$\rho_f =$ freshwater density (kg $\textrm{m}^{-3}$)

$\rho_s =$ saltwater density (kg $\textrm{m}^{-3}$)

$\bar{\rho}_\f =$ average density on freshwater side of the lock (kg $\textrm{m}^{-3}$)

$\rho^{\prime} =$ perturbation density (kg $\textrm{m}^{-3}$)

$\rho_{0} =$ reference density of water (kg $\textrm{m}^{-3}$)

$\sigma =$ density ratio between the salt and fresh water (--)

\bibliographystyle{apalike} 
\bibliography{bubble_curtain_dynamics_1}

@article{benjamin1968gravity,
  title={Gravity currents and related phenomena},
  author={Benjamin, T. B.},
  journal={Journal of Fluid Mechanics},
  volume={31},
  number={2},
  pages={209--248},
  year={1968},
  doi={10.1017/S0022112068000133},
  publisher={Cambridge University Press}
}

@article{shin2004gravity,
  title={Gravity currents produced by lock exchange},
  author={Shin, J. O. and Dalziel, S. B. and Linden, P. F.},
  journal={Journal of Fluid Mechanics},
  volume={521},
  pages={1--34},
  year={2004},
  doi={https://doi.org/10.1017/S002211200400165X},
  publisher={Cambridge University Press}
}

@article{kantarci2005bubble,
  title={Bubble column reactors},
  author={Kantarci, N. and Borak, F. and Ulgen, K. O.},
  journal={Process Biochemistry},
  volume={40},
  number={7},
  pages={2263--2283},
  year={2005},
  doi={https://doi.org/10.1016/j.procbio.2004.10.004},
  publisher={Elsevier}
}

@article{chisti1987airlift,
  title={Airlift reactors: characteristics, applications and design considerations},
  author={Chisti, M. Y. and Moo-Young, M.},
  journal={Chemical Engineering Communications},
  volume={60},
  number={1-6},
  pages={195--242},
  year={1987},
  doi={https://doi.org/10.1080/00986448708912017},
  publisher={Taylor \& Francis}
}

@article{schladow1993lake,
  title={Lake destratification by bubble-plume systems: Design methodology},
  author={Schladow, S. G.},
  journal={Journal of Hydraulic Engineering},
  volume={119},
  number={3},
  pages={350--368},
  year={1993},
  doi={https://doi.org/10.1061/(ASCE)0733-9429(1993)119:3(350)},
  publisher={American Society of Civil Engineers}
}

@article{wursig2000development,
  title={Development of an air bubble curtain to reduce underwater noise of percussive piling},
  author={W{\"u}rsig, B. and Greene Jr, C. R. and Jefferson, T. A.},
  journal={Marine Environment Research},
  volume={49},
  number={1},
  pages={79--93},
  year={2000},
  doi={https://doi.org/10.1016/S0141-1136(99)00050-1},
  publisher={Elsevier}
}

@techreport{abraham1973pneumatic,
  title={Pneumatic barriers to reduce salt intrusion through locks},
  author={Abraham, G. and Van der Burgh, P. and De Vos, P.},
  year={1973},
  institution={Rijkswaterstaat \& Delft Hydraulics Laboratory},
  url={https://resolver.tudelft.nl/uuid:e3e1ba53-8f09-43c1-8247-7c032b8604c2},
  address={Delft}
}

@article{fannelop1991surface,
  title={Surface current and recirculating cells generated by bubble curtains and jets},
  author={Fannel{\o}p, T. K. and Hirschberg, S. and K{\"u}ffer, J.},
  journal={Journal of Fluid Mechanics},
  volume={229},
  pages={629--657},
  year={1991},
  doi={https://doi.org/10.1017/S0022112091003208},
  publisher={Cambridge University Press}
}

@article{paillat2014entrainment,
  title={Entrainment in plane turbulent pure plumes},
  author={Paillat, S. and Kaminski, E.},
  journal={Journal of Fluid Mechanics},
  volume={755},
  pages={R2},
  year={2014},
  doi={https://doi.org/10.1017/jfm.2014.424},
  publisher={Cambridge University Press}
}

@article{o2024effect,
  title={The effect of bubble size on lock-exchange density currents through bubble screens},
  author={O’Mahoney, T. S. D. and Oldenziel, G. and Van Der Ven, P.},
  journal={Journal of Hydraulic Engineering},
  volume={150},
  number={3},
  pages={04024006},
  year={2024},
  doi={https://doi.org/10.1061/JHEND8.HYENG-13531},
  publisher={American Society of Civil Engineers}
}

@inproceedings{van2018methods,
  title={Methods to assess bubble screens applied to mitigate salt intrusion through locks},
  author={Van der Ven, P. P. D. and O’Mahoney, T. S. D. and Weiler, O. M.},
  booktitle={PIANC-World Congress Panama City. Panama City},
  pages={1--7},
  doi={},
  year={2018}
}

@article{spalding1971concentration,
  title={Concentration fluctuations in a round turbulent free jet},
  author={Spalding, D. B.},
  journal={Chemical Engineering Science},
  volume={26},
  number={1},
  pages={95--107},
  year={1971},
  doi={https://doi.org/10.1016/0009-2509(71)86083-9},
  publisher={Elsevier}
}

@article{cutroneo2014check,
  title={A check on the efficiency of an air-bubble screen using acoustic measurements and an artificial tracer},
  author={Cutroneo, L. and van der Goot, F. and Roels, A. and Castellano, M. and Radermacher, M. and Tucci, S. and Povero, P. and Canepa, G. and Capello, M.},
  journal={Journal of Soils and Sediments},
  volume={14},
  pages={1626--1637},
  year={2014},
  doi={https://doi.org/10.1007/s11368-014-0915-3},
  publisher={Springer}
}

@article{wang2024particle,
  title={Particle dynamic behaviors in bubble curtain barriers blocking the diffusion of sediment},
  author={Wang, Y. and Wu, J. and Shui, B. and Yang, J. and Wei, W.},
  journal={European Journal of Mechanics-B/Fluids},
  volume={106},
  pages={214--226},
  year={2024},
  doi={https://doi.org/10.1016/j.euromechflu.2024.04.013},
  publisher={Elsevier}
}

@book{abraham1962reduction,
  title={Reduction of salt water intrusion through locks by pneumatic barriers},
  author={Abraham, G. and Van der Burgh, P.},
  year={1962},
  publisher={Rijkswaterstaat \& Delft Hydraulics Laboratory},
  url={https://www.deltares.nl/expertise/publicaties/reduction-of-salt-water-intrusion-through-locks-by-pneumatic-barriers},
  address={Delft}
}

@article{bulson1961currents,
  title={Currents produced by an air curtain in deep water},
  author={Bulson, P. S.},
  journal={Dock Harbour Authority},
  volume={42},
  pages={15--22},
  year={1961},
  doi={}
}

@article{chen2014openfoam,
  title={OpenFOAM for computational fluid dynamics},
  author={Chen, G. and Xiong, Q. and Morris, P. J and Paterson, E. G. and Sergeev, A. and Wang, Y.},
  journal={Notices Of The American Mathematical Society},
  volume={61},
  number={4},
  pages={354--363},
  year={2014},
  doi={http://dx.doi.org/10.1090/noti1095}
}

@article{schiller1933drag,
  title={A drag coefficient correlation},
  author={Schiller, L.},
  journal={Zeit. Ver. Deutsch. Ing.},
  volume={77},
  pages={318--320},
  year={1933},
  doi={}
}

@article{tomiyama2002transverse,
  title={Transverse migration of single bubbles in simple shear flows},
  author={Tomiyama, A. and Tamai, H. and Zun, I. and Hosokawa, S.},
  journal={Chemical Engineering Science},
  volume={57},
  number={11},
  pages={1849--1858},
  year={2002},
  doi={https://doi.org/10.1016/S0009-2509(02)00085-4},
  publisher={Elsevier}
}

@article{drew1983mathematical,
  title={Mathematical modeling of two-phase flow},
  author={Drew, D. A.},
  journal={Annual Review of Fluid Mechanics},
  volume={15},
  number={1},
  pages={261--291},
  year={1983},
  doi={https://doi.org/10.1146/annurev.fl.15.010183.001401},
  publisher={Annual Reviews 4139 El Camino Way, PO Box 10139, Palo Alto, CA 94303-0139, USA}
}

@book{ishii2010thermo,
  title={Thermo-fluid dynamics of two-phase flow},
  author={Ishii, M. and Hibiki, T.},
  year={2010},
  doi={https://doi.org/10.1007/978-1-4419-7985-8},
  publisher={Springer Science \& Business Media},
  address={New York}
}

@article{weller1998tensorial,
  title={A tensorial approach to computational continuum mechanics using object-oriented techniques},
  author={Weller, H. G. and Tabor, G. and Jasak, H. and Fureby, C.},
  journal={Computers in physics},
  volume={12},
  number={6},
  pages={620--631},
  year={1998},
  doi={https://doi.org/10.1063/1.168744},
  publisher={American Institute of Physics}
}

@inproceedings{jasak2007openfoam,
  title={OpenFOAM: A C++ library for complex physics simulations},
  author={Jasak, H. and Jemcov, A. and Tukovic, Z.},
  booktitle={International workshop on coupled methods in numerical dynamics},
  volume={1000},
  pages={1--20},
  url={},
  year={2007}
}

@book{boussinesq1903theorie,
  title={Th{\'e}orie analytique de la chaleur: mise en harmonie avec la thermodynamique et avec la th{\'e}orie m{\'e}canique de la lumi{\`e}re},
  author={Boussinesq, J.},
  year={1903},
  address={Paris},
  doi={},
  publisher={Gauthier-Villars},
}

@book{nieuwstadt2016introduction,
  title={Introduction to Theory and Applications of Turbulent Flows},
  author={Nieuwstadt, F. T. M. and Westerweel, J. and Boersma, B.},
  year={2016},
  publisher={Springer},
  doi={https://doi.org/10.1007/978-3-319-31599-7},
  address={Cham}
}

@article{bacot2022bubble,
  title={Bubble curtains used as barriers across horizontal density stratifications},
  author={Bacot, A. and Frank, D. and Linden, P. F.},
  journal={Journal of Fluid Mechanics},
  volume={941},
  pages={A1},
  year={2022},
  doi={https://doi.org/10.1017/jfm.2022.142},
  publisher={Cambridge University Press}
}

@article{oldeman2020numerical,
  title={Numerical study of bubble screens for mitigating salt intrusion in sea locks},
  author={Oldeman, A. M. and Kamath, S. and Masterov, M. V. and O’Mahoney, T. S. D. and van Heijst, G. J. F. and Kuipers, J. A. M. and Buist, K. A.},
  journal={International Journal of Multiphase Flow},
  volume={129},
  pages={103321},
  year={2020},
  doi={https://doi.org/10.1016/j.ijmultiphaseflow.2020.103321},
  publisher={Elsevier}
}

@article{keetels2011field,
  title={Field study and supporting analysis of air curtains and other measures to reduce salinity transport through shipping locks},
  author={Keetels, G. and Uittenbogaard, R. and Cornelisse, J. and Villars, N. and van Pagee, H.},
  journal={Irrigation and Drainage},
  volume={60},
  pages={42--50},
  year={2011},
  doi={https://doi.org/10.1002/ird.679},
  publisher={Wiley Online Library}
}

@article{wen1987aeration,
  title={Aeration-induced circulation from line sources. I: Channel flows},
  author={Wen, J. and Torrest, R. S.},
  journal={Journal of Environmental Engineering},
  volume={113},
  number={1},
  pages={82--98},
  year={1987},
  doi={https://doi.org/10.1061/(ASCE)0733-9372(1987)113:1(82)},
  publisher={American Society of Civil Engineers}
}

@article{Li2025,
author = {Li, M. and Najjar, R. G. and Kaushal, S. and Mejia, A. and Chant, R. J. and Ralston, D. K. and Burchard, H. and Hadjimichael, A. and Lassiter, A. and Wang, X.},
title = {The Emerging Global Threat of Salt Contamination of Water Supplies in Tidal Rivers},
journal = {Environmental Science \& Technology Letters},
year = {2025},
volume = {},
pages = {available online 10.1021/acs.estlett.5c00505},
doi = {10.1021/acs.estlett.5c00505},
}

@article{kobus1968analysis,
  title={Analysis of the flow induced by air-bubble systems},
  author={Kobus, H. E.},
  journal={Coastal Engineering 1968},
  pages={1016--1031},
  url={https://ascelibrary.org/doi/pdf/10.1061/9780872620131.065},
  year={1968}
}

@article{ditmars1974analysis,
author = {Ditmars, J. D. and Cederwall, K.},
title = {Analysis of Air-Bubble Plumes},
journal = {Coastal Engineering},
chapter = {},
pages = {2209-2226},
doi = {10.1061/9780872621138.131},
URL = {https://ascelibrary.org/doi/abs/10.1061/9780872621138.131},
eprint = {https://ascelibrary.org/doi/pdf/10.1061/9780872621138.131},
year={1974}
}

@article{murai2025density,
  title={Density destratification by a single bubble plume in long horizontal fluid layers and in a dam lake},
  author={Murai, Y. and Tasaka, Y. and Noto, D. and Ulloa, H.},
  journal={Environmental Fluid Mechanics},
  volume={25},
  number={1},
  pages={4},
  year={2025},
  doi={https://doi.org/10.1007/s10652-025-10015-7},
  publisher={Springer}
}

@article{li2025numerical,
  title={A numerical study on temperature destratification induced by bubble plumes in idealized reservoirs},
  author={Li, Y. and Liu, D.},
  journal={Environmental Fluid Mechanics},
  volume={25},
  number={5},
  pages={41},
  year={2025},
  doi={https://doi.org/10.1007/s10652-025-10053-1},
  publisher={Springer}
}

@article{dugue2015influencing,
  title={Influencing flow patterns and bed morphology in open channels and rivers by means of an air-bubble screen},
  author={Dugu{\'e}, V. and Blanckaert, K. and Chen, Q. and Schleiss, A. J.},
  journal={Journal of Hydraulic Engineering},
  volume={141},
  number={2},
  pages={04014070},
  year={2015},
  doi={https://doi.org/10.1061/(ASCE)HY.1943-7900.0000946},
  publisher={American Society of Civil Engineers}
}

@article{covarrubias2025interaction,
  title={Interaction between a bubble curtain with waves and currents: implications on sediment dispersal},
  author={Covarrubias-Contreras, B. R. and Torres-Freyermuth, A. and Tinoco, R. O and Figueroa-Espinoza, B.},
  journal={Environmental Fluid Mechanics},
  volume={25},
  number={2},
  pages={14},
  year={2025},
  doi={https://doi.org/10.1007/s10652-025-10029-1},
  publisher={Springer}
}

@article{liu2025research,
  title={Research on the oil retention effect of pneumatic oil barriers under current and wave action},
  author={Liu, H. and Sun, H. and Jiao, B. and Lin, H. and Wang, G.},
  journal={PLoS One},
  volume={20},
  number={5},
  pages={e0322390},
  year={2025},
  doi={https://doi.org/10.1371/journal.pone.0322390},
  publisher={Public Library of Science San Francisco, CA USA}
}

@article{raaghav2025bubble,
  title={Bubble curtains in a lock-exchange flow: the importance of transient dynamics in the curtain-driven regime},
  author={Raaghav, S. K. R. and Driessen, R. J. A. and O'Mahoney, T. S. D. and Uittenbogaard, R. E. and Clercx, H. J. H and Duran-Matute, M.},
  journal={arXiv preprint arXiv:2511.23421},
  doi={https://doi.org/10.48550/arXiv.2511.23421},
  year={2025}
}

@article{prasad2026controlling,
  title={Controlling invasive carp ichthyoplankton dispersion using a streamwise-oriented bubble screen: A proof-of-concept validation in a laboratory flume},
  author={Prasad, V. and Doyle, H. F. and Suski, C. D. and Jackson, P. R. and George, A. E. and Fischer, J. R. and Stahlschmidt, B. H. and Herndon, A. M. and Tinoco, R. O.},
  journal={Journal of Great Lakes Research},
  pages={102784},
  year={2026},
  doi={https://doi.org/10.1016/j.jglr.2026.102784},
  publisher={Elsevier}
}

@article{weiler2026quantification,
  title={Quantification of Salt Intrusion caused by Navigation Locks and their Operation for Policy Analysis, Water Management or Salt Dispersion Modelling},
  author={Weiler, O. and Vreeken, T. and Maijvis, S. and Zuiderwijk, N. and O'Mahoney, T.},
  journal={Journal of Coastal and Hydraulic Structures},
  volume={6},
  doi={https://doi.org/10.59490/jchs.2026.0052},
  year={2026}
}

@article{bakker2026accurately,
  title={Accurately forecasting saltwater intrusion through navigation locks requires nautical traffic simulation modelling},
  author={Bakker, F. P and de Ruiter, J. W. E and Van der Hout, A. J. and Van Koningsveld, M.},
  journal={Ocean Engineering},
  volume={355},
  pages={124918},
  year={2026},
  doi={https://doi.org/10.1016/j.oceaneng.2026.124918},
  publisher={Elsevier}
}

@article{brevik2002flow,
  title={The flow in and around air-bubble plumes},
  author={Brevik, I. and Kristiansen, {\O}},
  journal={International journal of multiphase flow},
  volume={28},
  number={4},
  pages={617--634},
  year={2002},
  doi={10.1016/S0301-9322(01)00077-5},
  publisher={Elsevier}
}

@article{dissanayake2018integral,
  title={Integral models for bubble, droplet, and multiphase plume dynamics in stratification and crossflow},
  author={Dissanayake, A. L. and Gros, J. and Socolofsky, S. A},
  journal={Environmental Fluid Mechanics},
  volume={18},
  number={5},
  pages={1167--1202},
  year={2018},
  doi={https://doi.org/10.1007/s10652-018-9591-y},
  publisher={Springer}
}

@article{boufadel2020review,
  title={A review on multiphase underwater jets and plumes: Droplets, hydrodynamics, and chemistry},
  author={Boufadel, M. C. and Socolofsky, S. and Katz, J. and Yang, D. and Daskiran, C. and Dewar, W.},
  journal={Reviews of Geophysics},
  volume={58},
  number={3},
  pages={e2020RG000703},
  year={2020},
  doi={https://doi.org/10.1029/2020RG000703},
  publisher={Wiley Online Library}
}

@article{lee2024increasing,
  title={Increasing risks of extreme salt intrusion events across European estuaries in a warming climate},
  author={Lee, J. and Biemond, B. and de Swart, H. and Dijkstra, H. A.},
  journal={Communications Earth \& Environment},
  volume={5},
  number={1},
  pages={60},
  year={2024},
  doi={https://doi.org/10.1038/s43247-024-01225-w},
  publisher={Nature Publishing Group UK London}
}

@article{mcclimans2013pneumatic,
  title={Pneumatic oil barriers: The promise of area bubble plumes},
  author={McClimans, T. and Leifer, I. and Gj{\o}sund, S. H. and Grimaldo, E. and Daling, P. and Leirvik, F.},
  journal={Proceedings of the Institution of Mechanical Engineers, Part M: Journal of Engineering for the Maritime Environment},
  volume={227},
  number={1},
  pages={22--38},
  year={2013},
  doi={10.1177/1475090212450273},
  publisher={SAGE Publications Sage UK: London, England}
}

@article{yang2016large,
  title={Large-eddy simulation and parameterization of buoyant plume dynamics in stratified flow},
  author={Yang, D. and Chen, B. and Socolofsky, s. A. and Chamecki, M. and Meneveau, C.},
  journal={Journal of Fluid Mechanics},
  volume={794},
  pages={798--833},
  year={2016},
  doi={https://doi.org/10.1017/jfm.2016.191},
  publisher={Cambridge University Press}
}

@article{socolofsky2003liquid,
  title={Liquid volume fluxes in stratified multiphase plumes},
  author={Socolofsky, S. A. and Adams, E. E.},
  journal={Journal of Hydraulic Engineering},
  volume={129},
  number={11},
  pages={905--914},
  year={2003},
  doi={https://doi.org/10.1061/(ASCE)0733-9429(2003)129:11(905)},
  publisher={American Society of Civil Engineers}
}

@article{socolofsky2005role,
  title={Role of slip velocity in the behavior of stratified multiphase plumes},
  author={Socolofsky, S. A. and Adams, E. E.},
  journal={Journal of Hydraulic Engineering},
  volume={131},
  number={4},
  pages={273--282},
  year={2005},
  doi={https://doi.org/10.1061/(ASCE)0733-9429(2005)131:4(273)},
  publisher={American Society of Civil Engineers}
}

@article{lee2026comparative,
  title={Comparative morphological analysis between bubble curtains and bubble plumes},
  author={Lee, D. and Park, H.},
  journal={International Journal of Multiphase Flow},
  pages={105734},
  year={2026},
  doi={https://doi.org/10.1016/j.ijmultiphaseflow.2026.105734},
  publisher={Elsevier}
}

@article{beelen2024planar,
  title={Planar bubble plumes from an array of nozzles: Measurements and modelling},
  author={Beelen, S. and Krug, D.},
  journal={International Journal of Multiphase Flow},
  volume={174},
  pages={104752},
  year={2024},
  doi={https://doi.org/10.1016/j.ijmultiphaseflow.2024.104752},
  publisher={Elsevier}
}

@article{fraga2016influence,
  title={Influence of bubble size, diffuser width, and flow rate on the integral behavior of bubble plumes},
  author={Fraga, B. and Stoesser, T.},
  journal={Journal of Geophysical Research: Oceans},
  volume={121},
  number={6},
  pages={3887--3904},
  year={2016},
  doi={https://doi.org/10.1002/2015JC011381},
  publisher={Wiley Online Library}
}

@article{bombardelli2007modeling,
  title={Modeling and scaling of aeration bubble plumes: A two-phase flow analysis},
  author={Bombardelli, F. A. and Buscaglia, G. C. and Rehmann, C. R. and Rinc{\'o}n, L. E. and Garc{\'\i}a, M. H.},
  journal={Journal of hydraulic research},
  volume={45},
  number={5},
  pages={617--630},
  year={2007},
  doi={https://doi.org/10.1080/00221686.2007.9521798},
  publisher={Taylor \& Francis}
}

@article{nakai2002experimental,
  title={An experimental study on prevention of saline wedge intrusion by an air curtain in rivers},
  author={Nakai, M. and Arita, M.},
  journal={Journal of Hydraulic Research},
  volume={40},
  number={3},
  pages={333--339},
  year={2002},
  doi={https://doi.org/10.1080/00221680209499947},
  publisher={Taylor \& Francis}
}

@article{van1968prediction,
  title={Prediction of the extent of saltwater intrusion into estuaries and seas},
  author={Van Der Burgh, P},
  journal={Journal of Hydraulic Research},
  volume={6},
  number={4},
  pages={267--288},
  year={1968},
  doi={https://doi.org/10.1080/00221686809500241},
  publisher={Taylor \& Francis}
}

@article{socolofsky2002multi,
  title={Multi-phase plumes in uniform and stratified crossflow},
  author={Socolofsky, SA and Adams, EE},
  journal={Journal of Hydraulic Research},
  volume={40},
  number={6},
  pages={661--672},
  year={2002},
  doi={https://doi.org/10.1080/00221680209499913},
  publisher={Taylor \& Francis}
}

@article{dissanayake2021bubble,
  title={Bubble plume integral model for line-source diffusers in ambient stratification},
  author={Dissanayake, AL and Rezvani, M and Socolofsky, SA and Bierlein, KA and Little, JC},
  journal={Journal of Hydraulic Engineering},
  volume={147},
  number={5},
  pages={04021015},
  year={2021},
  doi={https://doi.org/10.1061/(ASCE)HY.1943-7900.0001885},
  publisher={American Society of Civil Engineers}
}

@article{Wuest1992,
author = {Wüest, Alfred and Brooks, Norman H. and Imboden, Dieter M.},
title = {Bubble plume modeling for lake restoration},
journal = {Water Resources Research},
volume = {28},
number = {12},
pages = {3235-3250},
doi = {https://doi.org/10.1029/92WR01681},
url = {https://agupubs.onlinelibrary.wiley.com/doi/abs/10.1029/92WR01681},
eprint = {https://agupubs.onlinelibrary.wiley.com/doi/pdf/10.1029/92WR01681},
year = {1992}
}

\end{document}